%% file: main.tex
\documentclass[acmsmall,screen,table,nonacm]{acmart}

\usepackage{balance}
\usepackage{amsmath}
\usepackage{xspace}
\usepackage{xcolor}
\usepackage[capitalise]{cleveref}
\usepackage[inline]{enumitem} 
\usepackage{graphicx}
\usepackage{pifont}
\usepackage{listings}

\usepackage{tikz}
\usepackage{tikz-qtree}

\usepackage[normalem]{ulem}

\usepackage{array,multirow,hhline}

\usetikzlibrary{trees}

\tikzset{
  solid node/.style={circle,draw,inner sep=1.2,fill=black},
  hollow node/.style={circle,draw,inner sep=1.2},
}

\newcommand{\healthy}{\ensuremath{{\sf Healthy}}}
\newcommand{\unhealthy}{\ensuremath{{\sf Unhealthy}}}

\usetikzlibrary{positioning,arrows,fit}

\newcommand{\xmark}{\raisebox{0em}{\color{red}\ding{55}}}
\newcommand{\cmark}{\raisebox{0em}{\color{green!70!black}\ding{51}}}

\newcommand{\fw}{\textsc{Catt}\xspace}

\usepackage{thmtools}

\definecolor{Gray}{gray}{0.88}
\definecolor{LightGray}{gray}{0.98}
\definecolor{DarkGray}{gray}{0.6}

\theoremstyle{definition}
\newtheorem{note}{Note}

\renewcommand{\paragraph}[1]{\smallskip \textbf{\emph{#1.}}}

\newtheorem{definition}{Definition}

\crefname{definition}{Def.}{Defs.}

\crefformat{section}{\S#2#1#3} % see manual of cleveref, section 8.2.1
\crefformat{subsection}{\S#2#1#3}
\crefformat{subsubsection}{\S#2#1#3}

\crefname{enumi}{}{}
\Crefname{enumi}{}{}

\newcommand{\hb}{\textit{D-HB}\xspace} %heartbeat
\newcommand{\su}{\textit{D-SU}\xspace} %secret update
\newcommand{\att}{\textit{D-ATT}\xspace} %attestation
\newcommand{\skupd}{\su\xspace}%\textit{SK-UPD}} %Session key update
\newcommand{\uk}{\textit{M-UK}\xspace} %Unique Keys
\newcommand{\nupd}{\su\xspace}%\textit{N-UPD}} %Nonce update
\newcommand{\puf}{\textit{M-PUF}\xspace} %Dynamic token
\newcommand{\attacktime}{\ensuremath{\delta_{PI}}\xspace} %Physical Intrusive attack time

\newcommand{\va}{\ia}

\newcommand{\ia}{\textit{IA}\xspace}

\newcommand{\microvisor}{\text{S$\mu$V}\xspace}

\newcommand{\attstart}{\#{\tt att\_start}\xspace}
\newcommand{\attend}{\#{\tt att\_end}\xspace}

\newcommand{\seda}{\text{SEDA}\xspace}
\newcommand{\darpa}{\text{DARPA}\xspace}
\newcommand{\sana}{\text{SANA}\xspace}
\newcommand{\lisa}{\text{LISA}\xspace}
\newcommand{\lisas}{\text{LISA-s}\xspace}
\newcommand{\lisaalpha}{\text{LISA-$\alpha$}\xspace}
\newcommand{\seed}{\text{SeED}\xspace}
\newcommand{\scapi}{\text{SCAPI}\xspace}
\newcommand{\pads}{\text{PADS}\xspace}
\newcommand{\erasmus}{\text{ERASMUS}\xspace}
\newcommand{\erasmusod}{\text{ERASMUS+OD}\xspace}
\newcommand{\salad}{\text{SALAD}\xspace}
\newcommand{\wise}{\text{WISE}\xspace}
\newcommand{\slimiot}{\text{slimIoT}\xspace}
\newcommand{\usaid}{\text{US-AID}\xspace}
\newcommand{\attauth}{\text{ATT-Auth}\xspace}
\newcommand{\museda}{\text{$\mu$SEDA}\xspace}

\newcommand{\sap}{\text{SAP}\xspace}
\newcommand{\mtra}{\text{MTRA}\xspace}
\newcommand{\radis}{\text{RADIS}\xspace}
\newcommand{\esdra}{\text{ESDRA}\xspace}
\newcommand{\eapa}{\text{EAPA}\xspace}
\newcommand{\shela}{\text{SHeLA}\xspace}
\newcommand{\healed}{\text{HEALED}\xspace}
\newcommand{\sadan}{\text{SADAN}\xspace}
\newcommand{\sfs}{\text{SFS}\xspace}
\newcommand{\dads}{\text{DADS}\xspace}
\newcommand{\pasta}{\text{PASTA}\xspace}
\newcommand{\sara}{\text{SARA}\xspace}
\newcommand{\simpleplus}{\text{SIMPLE+}\xspace}

\newcommand{\cora}{\text{CoRA}\xspace}

\newcommand{\fadia}{\text{FADIA}\xspace}

\newcommand{\rata}{\text{RATA}\xspace}

\newcommand{\bfb}{\text{BFB}\xspace}
\newcommand{\mton}{\text{M2N}\xspace}
\newcommand{\swarna}{\text{SWARNA}\xspace}

\newcommand{\swarnaagg}{\text{SWARNA-agg}\xspace}
\newcommand{\shots}{\text{SHOTS}\xspace}
\newcommand{\consensus}{\text{Poster}\xspace}
\newcommand{\mvs}{\text{MVS}\xspace}
\newcommand{\diat}{\text{DIAT}\xspace}

\newcommand{\hola}{\text{HolA}\xspace}

\newcommand{\fesa}{\text{FeSA}\xspace}

\newcommand{\scraps}{\text{SCRAPS}\xspace}

\newcommand{\tamarin}{\textsc{tamarin}\xspace}

\newcommand{\smart}{\text{SMART}\xspace}
\newcommand{\smartplus}{\text{SMART+}\xspace}
\newcommand{\trustlite}{\text{TrustLite}\xspace}
\newcommand{\tytan}{\text{TyTan}\xspace}
\newcommand{\hydra}{\text{HYDRA}\xspace}

\newcommand{\advsw}{${\it Adv}_{{\it SW}}$\xspace}
\newcommand{\advmsw}{$Adv_{{\it MSW}}$\xspace}  
\newcommand{\advpi}{${\it Adv}_{{\it PI}}$\xspace}
\newcommand{\advpni}{${\it Adv}_{{\it PNI}}$\xspace}

\newcommand{\advc}{${\it Adv}_{{\it DY}}$\xspace}

\newcommand{\etal}{\text{et al.}\xspace}
\begin{document}

\title[On the Design and Security of CRA Protocols]{On the Design and Security of  Collective Remote
  Attestation Protocols}

\author{Sharar Ahmadi}
\email{b.dongol@surrey.ac.uk}
\orcid{}
\affiliation{%
  \institution{University of Surrey}
  \city{Guildford}
  \country{UK}
}

\author{Jay Le-Papin}
\email{jay.le-papin@surrey.ac.uk}
\orcid{}
\affiliation{%
  \institution{University of Surrey}
  \city{Guildford}
  \country{UK}
}

\author{Liqun Chen}
\email{liqun.chen@surrey.ac.uk}
\orcid{0000-0003-0446-3507}
\affiliation{%
  \institution{University of Surrey}
  \city{Guildford}
  \country{UK}
}

\author{Brijesh Dongol}
\email{b.dongol@surrey.ac.uk}
\orcid{0000-0003-0446-3507}
\affiliation{%
  \institution{University of Surrey}
  \city{Guildford}
  \country{UK}
}

\author{Sasa Radomirovic}
\email{s.radomirovic@surrey.ac.uk}
\orcid{}
\affiliation{%
  \institution{University of Surrey}
  \city{Guildford}
  \country{UK}
}

\author{Helen Treharne}
\email{h.treharne@surrey.ac.uk}
\orcid{}
\affiliation{%
  \institution{University of Surrey}
  \city{Guildford}
  \country{UK}
}

% Submissions should be anonymized. See the CFP for details on how to anonymize your paper, including any references to your own work.
%\author{\IEEEauthorblockN{Sharar Ahmadi, Jay Le-Papin, Liqun Chen, Brijesh Dongol, Sasa Radomirovic and Helen Treharne}
%\author{\IEEEauthorblockN{Anonymous Authors}
%  \IEEEauthorblockA{% Department of Computer Science \\
%    University of Surrey\\
%  Guildford, UK }
%}

% The author information is skipped here, but can be used to include author information in the publication.
%\iffalse
% \author{\IEEEauthorblockN{1\textsuperscript{st} Given Names Surname}
% \IEEEauthorblockA{\textit{University of Surrey} \\
% Guildford, UK\\
% email address or website URL}
% \and
% \IEEEauthorblockN{2\textsuperscript{nd} Given Names Surname}
% \IEEEauthorblockA{\textit{University of Surrey} \\
% Guildford, UK \\
% email address or website URL}
% \and
% \IEEEauthorblockN{3\textsuperscript{rd} Given Names Surname}
% \IEEEauthorblockA{\textit{University of Surrey} \\
% Guildford, UK \\
% email address or website URL}
% \and
% \IEEEauthorblockN{4\textsuperscript{th} Given Names Surname}
% \IEEEauthorblockA{\textit{University of Surrey} \\
% Guildford, UK \\
% email address or website URL}
% \and
% \IEEEauthorblockN{5\textsuperscript{th} Given Names Surname}
% \IEEEauthorblockA{\textit{University of Surrey} \\
% Guildford, UK \\
% email address or website URL}
% %% IEEE format can accommodate up to six authors this way
% }
% %\fi

\begin{abstract}
% The pervasiveness of Internet of Things (IoT) has meant that
 %  With the growth of the Internet of Things (IoT), millions of
 %  low-power % smart
 %  devices with limited computing resources are now connected to form
 %  heterogeneous self-organizing networks called \emph{swarms}.
 % % \sr{Are there obvious examples of such networks?}
 %  % These devices
 %  % typically, have limited computing resources,
 %  %%% SR: Sentence below says limited comp power => target for attack
 %  %%%     I don't think that's necessarily true. So it would require
 %  %%%     evidence. Hence removed.
 %  %Unfortunately, the limitations on computing power makes them a
 %  %target for attacks, with the potential of disrupting the entire
 %  %network. In this context,
 %  These networked devices are a natural target for attacks and must
 %  therefore be monitored.
  \emph{Collective remote attestation} (CRA)
  is a security service
%\underline{widely used security service}
%\sr{We need evidence for ``widely used''} 
  that aims to efficiently identify
  compromised % (and potentially malicious)
  (often low-powered) devices in a (heterogeneous)
  network. %, called \emph{provers}.
  % CRA protocols have been designed for a wide range of application
  % domains, and hence, must protect against a range of different
  % adversaries by providing some level of security.
  The last few years have seen an extensive growth in CRA protocol
  proposals, showing a variety of designs guided by different network
  topologies, hardware assumptions and other functional requirements.
  However, they differ in their trust assumptions, adversary models
  and role descriptions making it difficult to uniformly assess their
 security guarantees. % \SR{The last two sentences both say new
 %    different things were developed, but the second collection of
 %    different things is bad. We need to tell the reader why it is bad.
 % For example: ``... '' }

%  Due to the vast design space, a systematic study
%  of the protocols is needed to categorise them and compare them
%  according to well-defined metrics.
  %In this paper, we study a wide
  %range of CRA protocols and define a set of fine-grained metrics to
  %better understand the design space.

  In this paper we present \fw, a unifying framework for CRA protocols that enables them to be compared systematically, based on a comprehensive study of 40 CRA protocols and their adversary models. 
  \fw characterises the roles that devices can take and 
  based on these we develop a novel set of security properties for CRA protocols.
  % defines notions of trustworthiness of attestation results.
  % In this paper, we present a review of CRA protocols, based on a comprehensive review of 40 protocols, and develop \fw, a unifying framework that enables the roles and notions of trustworthiness of CRA protocols to be compared systematically.
  % Which enables the roles and notinos of trustworthiness to be compared systematically
  % develop \fw, a unifying framework for CRA protocols.
  %  based on a comprehensive review of 40 protocols from the
  % literature. 
  % \fw includes a novel model of CRA protocols that
  % characterises the roles that devices can take and defines notions of
  % trustworthiness of attestation results. 
  % Furthermore, review and systematise mitigation of software attacks as well as detection techniques for physical attacks. 
  We then classify the security aims of all the studied protocols. 
  % We develop a novel set of a security properties for CRA protocols based on our unified model, which we use to
%  \sout{compare}
% classify the security aims of all of the protocols that we study.
%   \sa{
% In our comparison, we employ existing metrics from the literature and introduce new ones. We then use the protocols to systematise and formally define nine principal CRA
% %   security properties, which for the first time provide a way of comparing all CRA protocols in a unified way, and identify the appropriate properties for each of the reviewed CRA protocols.} 
% We encode a subset of our security properties in \tamarin and demonstrate their applicability.
We illustrate the applicability of our security properties by encoding them in the \tamarin prover and verifying the \simpleplus protocol against them.

% We validate a subset of our security properties by encoding them in the \tamarin prover and verifying the \simpleplus protocol against them.

% We validate a subset of our security properties by implementing them in a symbolic model of the \simpleplus protocol, which we verify using the Tamarin prover.
% that % We also categorise the
  % adversaries which the security properties are defined in their
  % presence.
%\sr{Please check: Is the underlined sentence in the abstract above
%  correct?}\sa{some have formal definitions, some have formal proofs, but they are not general ones, just ad hoc definitions and proofs.}
%\bd{I suggest deleting that sentence}
\end{abstract}

\maketitle

% we systematically compare the designs, adversary
%   models and attestation capabilities of 40 CRA protocols from the
%   literature. To this end, 

% For lemma figure
\lstset{%
  basicstyle=\footnotesize\ttfamily,
  frame=single,
  morecomment=[l][\color{green!70!black}\bf]{//},
  morecomment=[f][\color{green}][0]{*},
  % morecomment=[f][\color{red}][0]{\#},
  escapeinside={(*}{*)},
  % keywords={not,All,Ex,lemma,functions},
  keywordstyle = {\bfseries\color{blue}}
  }

% \begin{IEEEkeywords}
% collective remote attestation, security properties
% \end{IEEEkeywords}

\input{intro}

\input{slim-overview}

\input{adversary_model}
\input{IA}
\input{security_properties}

\input{validation}

% \input{relatedwork}

\input{conclusion}
\section*{Acknowledgements}
\label{sec:ack}
This work is supported by VeTSS and EPSRC grants EP/Y036425/1,
EP/X037142/1, EP/X015149/1, EP/V038915/1 and EP/R025134/2.
%\newpage

% \input{related_works}

%\input{example}

% \balance

% \bstctlcite{IEEEexample:BSTcontrol}
\bibliographystyle{Acm-Reference-Format}
\bibliography{references}

\appendix

 \input{appendix}

\end{document}

%% file: intro.tex
%!TeX root = main.tex

\section{Introduction}\label{sec:intro}
An increasing number of heterogeneous % interconnected
% embedded IoT
devices are deployed in distributed and autonomous networks with
applications including % smart
medical systems~\cite{smart-medical-systems} to % , smart
% cities~\cite{privhome}, farming~\cite{smart-farming} and
robotics~\cite{uav}.
% and environment monitoring~\cite{environmental-monitoring}.
% , smart
% parking system~\cite{smart-parking}, smart
% farming~\cite{smart-farming}, building
% automation~\cite{building-automation}, Unmanned aerial vehicles
% (UAVs)~\cite{}, ,
% transportation~\cite{survey-iot-transportation}, industrial control
% systems~\cite{industrial-control}, and safety-critical
% systems~\cite{survey-safety-critical-systems}.
%I am not sure about the contents of ref. oil-gas
% This is referred to as Internet of Things (IoT), where physical systems are connected over network to exchange information among themselves and with their 
% surroundings~\cite{survey-iot,IoT-survey}. 
These pervasive devices often operate in large, dynamic, and
self-organising networks, called \emph{swarms}~\cite{seda}. The devices themselves
are typically low cost and have limited memory and processing
capabilities. This unfortunately, makes them targets for a wide range
of % cyber-
attacks~\cite{demystifying, survey-security-privacy, kuang2022survey, hassija2019survey}. 

%\cite{survey-iot-practical}.

One way to monitor these attacks is {\em remote attestation}
(RA)~\cite{steiner,state-of-the-art,survey-device-attestation} whereby
a device, called a \emph{verifier}, checks the software state of a
remote device, called a \emph{prover}. In RA protocols, a prover sends
a report containing a measurement (e.g., of its software state) to the
verifier, which can be matched against expected values to determine
whether the prover can be trusted.  However, when attesting multiple
provers, an RA protocol must be executed for each prover individually,
which is not scalable. Therefore, \emph{collective remote attestation}
(CRA) protocols (aka {\em swarm attestation} protocols) have been
developed, which allow a verifier to efficiently attest a large number
of provers~\cite{seda,Ambrosin,kuang2022survey}. CRA protocols
can support {\em heterogeneous systems} (e.g., IoT
networks)
%systems), 
where
%different 
provers may have different hardware and software configurations (e.g.,~\cite{scapi}).
Moreover, the provers may not be stationary (e.g., as part of an
autonomous drone~\cite{DBLP:conf/wimob/ImamLA21}), which further increases the
design space for CRA protocols (e.g., as found in~\cite{salad}).

% \bd{Add something about devices moving around increasing the challenge}

Since the first CRA protocol, \seda~\cite{seda} was proposed, many new % a large number of
protocols have been developed %, covering a wide range of designs space,
to address different assumptions on interactivity, network topologies,
device mobility, verifier behaviour, and reporting and validation
requirements.
%In order to clarify the design aspects and key characteristics of the
%protocols,
The large number of protocols and their vast design space 
led to several efforts to classify the % key features of CRA
protocols.  % in a unifying framework.
% Some prior works, notably
Carpent \etal~\cite{lisa}, Ambrosin \etal~\cite{Ambrosin}, and Kuang
\etal~\cite{kuang2022survey} have compared CRA protocols with respect to 
operational assumptions and design features such as the % underlying
network's topology and dynamicity, hardware requirements and cost, and
quality of attestation.

\paragraph{Contributions}
%\sout{In this paper, we systematically review the adversary models, trust
%assumptions and security requirements of 40 CRA protocols, and related
%literature on RA protocols, which we used to develop a new unifying
%framework, \fw (\underline{C}ollective
%\underline{Att}estation). \smallskip}
% In this paper, w
We present \fw a new unifying framework for \underline{C}ollective
\underline{Att}estation. \fw is the result of our findings from 
our systematic review of the adversary models, trust
assumptions and security requirements of 40 CRA protocols and related
literature on RA protocols. 
%Specifically, 
Namely, we make the following contributions with \fw.% \smallskip

\begin{enumerate}[itemsep=2pt,leftmargin=0pt,itemindent=0pt]
\item [] (1) {\em Unified Protocol Model (\cref{sec:overview})}. The literature on CRA protocols
  (including the aforementioned
  reviews~\cite{Ambrosin,kuang2022survey}) has generally evolved from
  RA protocols. However, different authors have applied extensions to
  the RA terminology in an ad hoc (often inconsistent) manner. \fw
  includes a novel protocol model, building on the RATS
  RFC~\cite{10.17487/RFC9334}, that clarifies the existing terminology
  and defines new roles such as a \emph{relying party} and
  \emph{delegated trusted party} for CRA protocols. We show that the
  \fw protocol model serves as a framework within which all of the
  protocols that we study can be characterised.

\item [] (2) {\em Trust and Threat Models (\cref{sec:threats to
      integrity}).} We review the possible targets of attestation
  from which a measurement is generated
  and clarify how validation of
  measurements can be trusted.  Additionally, we take the existing
  threat models~\cite{seed,Ambrosin,Abera-Invited} and recast them
  within \fw.
\item [] (3) {\em Defence Mechanisms Against Software and Physical Attacks (\cref{protection and mitigation}).}
  We review techniques to defend against software attacks % , which
  % provide guarantees that are assumed by the \fw model.\SR{I don't
  %   understand ``assumed by the model''.}
  % Additionally,
  and \emph{systematise} the mitigation and detection techniques for
  physical attacks, condensing approaches into five broad categories.
\item [] (4) {\em Security Definitions (\cref{subsec:verifier
      authentication} and \cref{sec:security properties}).}  Unlike previous works, which have
  introduced ad hoc (informal) security definitions, \fw introduces a
  novel set of a security properties for CRA protocols based on our
  unified model, which we use to 
%\sout{compare}
  classify the security goals of all 40 of the protocols that we
  study.  The framework comprises one authentication property and
  eight attestation properties. The former is used by many protocols
  to prevent unauthorised parties from starting the costly attestation
  process. The latter properties capture the main security goals,
  enabling comparison of CRA protocols % and we
  % have designed them to enable the comparison of CRA protocols 
  across
  three dimensions: (i) accuracy with respect to the attestation
  target, (ii) granularity of the attestation result, and (iii)
  synchronicity of attestation reports.

  % \fw our review
  % and introduces a novel unified approach for comparing CRA protocols.

  % The literature currently does not
  % include a rigorously defined set of security properties for CRA
  % protocols. Using 

%   Based on our analysis, we provide the first framework that articulates
% formal security properties for CRA protocols. 
% \sa{} 
% in terms of
% \begin{enumerate*}[label={\bfseries\arabic*)}]
% \item {\em attestation properties}, namely, {\em individual
%     (a)synchronous weak/strong attestation} and {\em group
%     (a)synchronous weak/strong attestation}, and 
% \item \emph{initiator authentication}, which guarantees that whenever
%   a prover receives a message, the prover can decide whether that
%   message has been sent to it by an authorised party. % or not.
% \end{enumerate*}
% Our attestation properties

\item [] (5) {\em Property Encoding (\cref{sec:examples-validation}).} % We formalise \fw security properties in
  Finally, we illustrate the \fw security properties by showing that they can be encoded in the Tamarin prover and demonstrate which ones hold for one of the reviewed CRA protocols,
  SIMPLE+~\cite{simple}. The significance of the encoding is that
  the properties can then be used directly in formal analyses of other
  CRA protocols. 
  % We expect the remaining properties could be encoded in a similar manner. 
  % ... We
  % validate our security definitions by implementing them in a symbolic
  % model of the SIMPLE+ protocol, which we verify using the Tamarin
  % prover.
  % However, a comparison of CRA protocols with respect to their
  % security aims is missing, because there is no common framework of
  % rigorously defined security properties.

\end{enumerate}

%% file: slim-overview.tex
\newcommand{\pmf}{PM}
\newcommand{\pmr}{PM}
\section{A Unified Protocol Model for CRA}
\label{sec:overview}

% \jlp{This entire section needs to be rewritten to bring it in line with the RFC}

%\sr{Need to refer to Table}
%In this section, we define common terminology found in the literature
%that is used in CRA protocols (see~\cref{subsec:preliminaries}). 

%\subsection{Preliminaries}\label{subsec:preliminaries}

% \bd{Sharar, Jay - double check reference [11] - add others?}

In the literature, CRA is often considered to be a protocol between a
non-empty set of \emph{provers} and a {\em verifier} (or set of
verifiers)~\cite{seda,Ambrosin,kuang2022survey}.  However, in our
study, we discover that % turns out
% that
this formulation is not suitable for describing CRA protocols in
general, because the precise trust assumptions and functionality of a
verifier varies. This is unlike RA protocols, where a verifier is
generally assumed to be trusted and have a fixed role of deciding
whether the prover it interacts with is
healthy~\cite{10.17487/RFC9334}. CRA protocols follow a variety of
designs ranging from untrusted verifiers~\cite{fesa} to not using
verifiers at all~\cite{healed}. Even when trusted verifiers are
assumed, their functionalities often differ from verifiers in RA
protocols, e.g., verifiers in a CRA protocol may only collect results
generated by provers~\cite{simple}. % Overall this
% makes it difficult to define the security properties that generally
% apply to CRA protocols.

In \fw, we thus define a CRA protocol as one that specifies the
behaviour of a % non-empty
set of \emph{provers} and % possibly a number
% of
some intermediary devices to enable a {\em relying
  party}~\cite{DBLP:journals/ijisec/CokerGLHMORSSS11,10.17487/RFC9334}
to determine whether the provers are trustworthy.
%This may be achieved
%using a number of other intermediary parties, which we discuss in more
%detail below.
% An optional, but
% common role in CRA protocols is the \emph{aggregator}, which combines
% different reports into a single report.
% % Some CRA protocols refer to the relying party as the operator~\cite{scapi}
% % % (which collects the final attestation result from the
% % % verifier).  
% % or the network owner~\cite{sana}.  
% Some CRA protocols refer to the relying party as the
% operator~\cite{scapi} or the network owner~\cite{sana}.
%%%%% Measurement %%%%%%

The core of a CRA protocol is the {\it measurement}, which provides information
about the prover's state at some \emph{time}. The notion of time used
by protocols is typically a real-time clock~\cite{seed,pads,pasta} or
a sequence number~\cite{sana,simple,lisa}. The measurement is sent by
a prover in an \emph{attestation report}, usually signed or alongside
a MAC to authenticate the report. Attestation reports can be combined
to create an \emph{aggregated report}, and an aggregated report can be
combined with other attestation or aggregated reports to create a more
informative aggregated report. In the literature and also in this
paper, the term \emph{report} is overloaded to mean both a single
attestation report and an aggregated report.

A CRA protocol depends on a (potentially abstract) list of \emph{expected} states and a corresponding list of expected measurements, i.e., measurements that indicate the provers to be trustworthy.
% are considered to be acceptable.  %% SR changed to "trustworthy" to
% be consistent with our definition of CRA protocol above.
We say that a prover has a {\em healthy} (or
{uncompromised}~\cite{seda}) {\em status} at a particular time iff,
the prover's state at that time is in the list of acceptable states for that prover at that time. Conversely, a prover has an  {\em unhealthy} (or
{compromised}~\cite{seda}) {\em status} if the state is
not an expected state. %  \jlp{[Does this paragraph
  % belong here? Or do we leave discussion of unresponsive provers to
  % the Weak/Strong property?]} 
Depending on the CRA protocol, the
prover
% may be judged to be 
% unhealthy if it does not respond in a timely manner~\cite{seda,simple}.
might be considered by the relying party to be unhealthy if a response
is not received in a timely manner~\cite{seda,simple}, be that due to the
prover not completing the protocol, interference by an adversary, or
other reasons. Other protocols distinguish between unhealthy and
unresponsive~\cite{pads}.  The definition of the expected measurements
varies between protocols
%%%% SR: Not clear why Ambrosin is singled out here:
%% Ambrosin \etal~\cite{Ambrosin} define that a prover's measurement
%% is an expected one if the prover is running the latest legitimate software version.
%%%% SR: Could keep the Ambrosin example if we give at least one other example.
and many CRA schemes keep it abstract, which we
follow.
% The security
%   requirements can be applied to different CRAs whose failing
%   attestation may be a result of adversarial attacks or or network
%   delay, and who can be followed by further actions such as resetting
%   unhealthy provers (e.g., \healed~\cite{healed}) or disconnecting
%   them from the swarm (e.g., \cite{}).
%%%%% Roles %%%%%

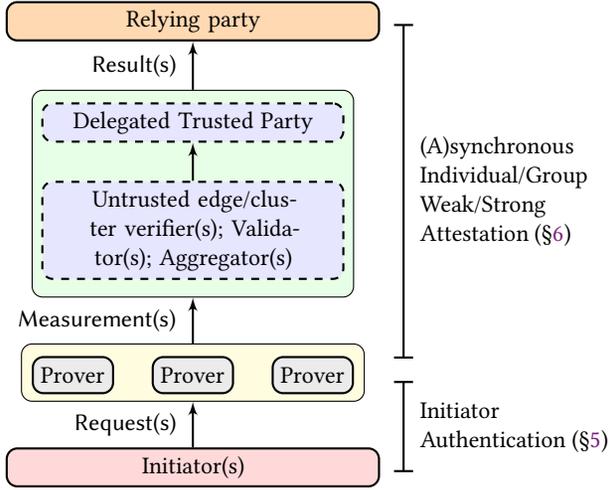
\begin{figure}[!t]
  \centering
  \small
  \scalebox{1}{
  \begin{tikzpicture}[
    box/.style={rectangle,draw,fill=DarkGray!20,node distance=1cm,text width=12em,text centered,rounded corners,minimum height=1.5em,thick},
    boxa/.style={rectangle,draw,fill=DarkGray!20,node distance=1cm,text centered,rounded corners,minimum height=1.5em,thick},
    arrow/.style={draw,-latex',thick}  ]

  \node [box,fill=orange!30, text width=15em] (relying) {Relying party};

  \node [below=0.8 of relying] (DTP) {\phantom{Delegated Trusted Party}};
  \node [below=0.5 of DTP] (untrusted) {\phantom{
      \begin{tabular}[t]{@{}l@{}}
        \ \ \ \ Untrusted edge/cluster \ \ \ \  \\ verifier(s) ;
        Validator(s); \\ Aggregator(s)
      \end{tabular}
    }};

    \node [rectangle,fill= green!10, draw=black,rounded corners, inner sep=0.5em,fit=(DTP) (untrusted)] (verification) {};
  \node [box,below=0.8 of relying,fill=blue!10,dashed] (DTP) {Delegated Trusted Party};
  \node [box,below=0.5 of DTP,fill=blue!10,dashed] (untrusted) {Untrusted edge/cluster verifier(s); Validator(s); Aggregator(s)};

  \node [below=1 of untrusted] (prover2) {\phantom{Prover}};
  \node [left=0.5 of prover2] (prover1) {\phantom{Prover}};
  \node [right=0.5 of prover2] (prover3) {\phantom{Prover}};

  \node [rectangle,,fill = yellow!15,draw=black,rounded corners, inner sep=0.5em,fit=(prover1) (prover3)] (provers) {};

  \node [boxa,below=1 of untrusted] (prover2) {Prover};
  \node [boxa,left=0.5 of prover2] (prover1) {Prover};
  \node [boxa,right=0.5 of prover2] (prover3) {Prover};

    \node [box,below=0.6 of provers, fill=red!15, text width=15em] (initiators) {Initiator(s)};

    % \node [above right=0.1 and 0.1 of verification, anchor=center] (verLabel) {Verification};

  \path [arrow] (initiators) -- (provers) node [midway,left=0.1] (TextNode) {\sf Request(s)};
  \path [arrow] (provers) -- (verification) node [midway,left=0.1] (TextNode) {\sf \small Measurement(s)};
  \path [arrow] (verification) -- (relying) node [midway,left=0.1] (TextNode) {\sf \small Result(s)};
  \path [arrow] (untrusted) -- (DTP) ;

  \node [below right=-0.2 and 0.2 of initiators] (ISP) {};
  \node [above=1.2 of ISP] (PSP) {};
  \node [above right=-0.3 and 0.2 of relying] (RSP) {};
  \node [below=4.4 of RSP] (RSP2) {};

  \path [|-|,draw,thick] (ISP) -- (PSP)
  node[midway,right=0.1] {
    \begin{tabular}[t]{@{}l@{}}
      Initiator \\Authentication (\cref{subsec:verifier authentication})
    \end{tabular}
} ;
  \path [|-|,draw,thick] (RSP) -- (RSP2)
  node[midway,right=0.1] {
    \begin{tabular}[t]{@{}l@{}}
      (A)synchronous \\
      Individual/Group \\
      Weak/Strong \\
      Attestation (\cref{sec:security properties})
    \end{tabular}
} ;
  
\end{tikzpicture}}
% \vspace{-5pt}
\caption{Summary of roles and security properties in the \fw model.
  The prover(s) (yellow box) generate measurement(s) in response to
  request(s) from the initiator(s). Measurement(s) are verified by a
  range of mechanisms (green box) to generate result(s) that are sent
  to the relying party. We used dashed boxes to indicate that there
  are several design choices when selecting components that generate
  result(s) from measurement(s). }
\label{fig:roles}
% \vspace{-10pt}
\end{figure}
%The condition of being healthy or unhealthy is a prover's
%\emph{status}.  The status observed by the verifier, the
%\emph{observed status}~\cite{Jay}, may be limited to just healthy or
%unhealthy~\cite{simple}, or may include other options such as
%\emph{unknown}~\cite{pads}.

%%%%% Trust %%%%%%

We assume that the attestation procedure is triggered by an
\emph{initiator} that signals provers to take their measurements for
validation. There are a wide range of options for initiating
attestation and the initiator and the provers are not always distinct
devices (see \cref{subsec:interactive} for details). Other common
roles in CRA protocols may include a \emph{verifier} (or sometimes
group of verifiers) that performs attestation on behalf of the relying
party; a {\em validator} (or sometimes group of validators) that
checks whether the measurement generated by the provers are healthy;
and an \emph{aggregator} (or sometimes group of aggregators) that
combines different reports into a single report to improve
scalability.

As we shall see in \cref{subsec:threats to verifier integrity}, a CRA
protocol may also use a trusted third party, which we refer to as a
\emph{delegated trusted party}, that may (but not always) include a
verifier. An overview of the roles is given in \cref{fig:roles}. This
use of a delegated trusted party means that our systematisation for
CRA protocols remains compatible with the roles assumed by RA
protocols~\cite{10.17487/RFC9334}.

As % we discuss in \cref{sec:security properties} and
highlighted in
\cref{fig:roles}, % in our systematisation, for the first time, we
% identify several
\fw includes several security properties, split into two types. The
first is an authentication property between initiators and
provers, and the second describes security between the provers and the relying party. % , and the % . %  and
% % % forms a hierarchy. 
% the second 

\section{Device Integrity}\label{sec:threats to integrity}

% \bd{What is the purpose of this section wrt \fw? Clarify what is
%   outside and what is inside the framework.}

The purpose of the measurement in a CRA protocol is to provide
evidence of the integrity of the prover's software. 
Different CRA protocols differ in their \emph{attestation targets} % ,
% i.e., the precise components of a prover that are untrusted and thus 
% examined in order to produce the measurement
(see \cref{subsec:threats to prover integrity}). % We describe these in
% .
% ; we refer to these
% different aspects as {\it attestation targets}. The literature does
% not provide a clear systematization of these kinds of attestations
% targets and here we provide the following classification (see column
% 11).
%
% In the previous section we introduced commonly accepted terminology
% and in this section we introduce new terminology that enables us to
% identify how measurements are achieved. We also examine the different
% threat models that can exist within CRA protocols.
%
% 
% In \cref,
We describe how trustworthiness of the verification (of the
measurement) is achieved in terms of the \fw model in
\cref{subsec:threats to verifier integrity}, and models of attacks and adversaries in %  % , due to the
% % fact that different roles in CRA protocols may be executed by the same
% % device.
% % when a verifier's judgement can be trusted.
% % Finally, CRA protocols also differ with respect to the threats to
% % device integrity and to the execution of the CRA protocol that they
% % consider. 
% We present the attacks and adversary models that have been
% considered 
\cref{subsec:threats to CRA integrity}.
These considerations lead to different protocol designs which we
summarise in \cref{table:categories}.
% We focus on aspects that affect
% the security properties that we develop in this paper.
% Other
% considerations, such as the topology and dynamicity of the network are
% presented in the Appendix. %\cref{app:more metrics}.
% % \bd{Add a reference to \cref{table:categories} and a short discussion.}

% In this section we systematically examine potential threats to devices

% (see~\cref{subsec:threats to prover integrity}), verifiers
% (see~\cref{subsec:threats to verifier integrity}), and CRA integrity
% (see).

\newcommand{\TopST}{S. Tree}
\newcommand{\TopAgg}{Agg. Layer}
\newcommand{\TopBBT}{BB. Tree}
\newcommand{\TopBroad}{Distributed}
\newcommand{\TopCluster}{Cluster}
\newcommand{\TopPubSub}{PubSub}
\newcommand{\TopHierarchical}{Hierarchical}
\setlength{\tabcolsep}{2pt}

\newcommand{\DynStatic}{Static}
\newcommand{\DynQStatic}{Q. Static}
\newcommand{\DynDynamic}{Dynamic}
\newcommand{\DynHDynamic}{H. Dynamic}

\begin{table*}[!t]
\footnotesize
\begin{center}
  \centering\caption{\label{table:categories} Summary of CRA protocol
    designs}
  \scalebox{0.95}{
\begin{tabular}{|p{0.16\linewidth}|p{0.04\linewidth} %p{0.07\linewidth}p{0.097\linewidth}p{0.085\linewidth}
  cccccc|}
  \hline 
  \rowcolor{DarkGray!70!white}
  {\bf Protocol} & {\bf Year} % & {\bf Topology} (\cref{subsec:topology}) & {\bf Dynamicity} (\cref{subsec:dynamicity}) &{\bf (De)Centrali-sation} (\cref{subsec:centralised})
&
  \begin{tabular}[t]{@{}l@{}}
    {\bf Attestation} \\
    {\bf Target} (\cref{subsec:threats to prover integrity})
               \end{tabular}

  &
    \begin{tabular}[t]{@{}l@{}}
      {\bf DTP }  (\cref{subsec:threats to verifier integrity})
    \end{tabular}
& \begin{tabular}[t]{@{}l@{}}
    {\bf Validator} \\(\cref{subsec:threats to verifier integrity})
  \end{tabular}
%& \begin{tabular}[t]{@{}l@{}}
%    {\bf Device} \\
%    {\bf Availability} (\cref{subsec:mitigation-detection-da})
%  \end{tabular}
&
 \begin{tabular}[t]{@{}l@{}}
   {\bf Interactivity} \\
   (\cref{subsec:interactive})
 \end{tabular}
&
\begin{tabular}[t]{@{}l@{}}
  {\bf Attestation} \\ {\bf Set} (\cref{subsec:attestation set})
\end{tabular}
&
  \begin{tabular}[t]{@{}l@{}}
    {\bf QoSA} \\ (\cref{sec:attest-prop-1})
  \end{tabular}

\\
  \hline 
  \seda\cite{seda} & 2015 & \pmf& % \TopST & \DynStatic & Centralised & 
                                                                Verifier & Self-L % & \xmark 
& Interactive-T  & All & Binary\\
%\hline
\rowcolor{Gray}
\darpa\cite{darpa}+\seda\cite{seda}& 2016 &\pmf& % \TopST & \DynStatic %dynamic 
% & Centralised & 
                Verifier & Self-L % & \cmark 
& Interactive-T  & All & Binary   \\
%\hline
\sana\cite{sana}& 2016 &\pmf& % \TopAgg & \DynStatic %dynamic 
% & Centralised & 
                Verifier & Self-V % & \xmark 
& Interactive-T  & All & List  \\
%\hline
\rowcolor{Gray}
\lisas\cite{lisa}& 2017 &\pmf& % \TopST & \DynQStatic &Centralised & 
                                                              Verifier & Self-L % & \xmark 
& Interactive-T  & All& List  \\
%\hline
\lisaalpha\cite{lisa}&2017 &\pmf& % \TopST & \DynQStatic &Centralised & 
                                                                 Verifier & Verifier % & \xmark 
& Interactive-T  & All & List \\
%\hline
\rowcolor{Gray} 
\seed\cite{seed}+\seda\cite{seda}& 2017 &\pmf&% 
                Verifier & (1) % & \xmark  & %Noninteractive-U  
& Noninteractive & All & Binary  \\ %DoS-V
%\hline
\scapi\cite{scapi} & 2017 &\pmf& 
                                 Verifier & Self-V % & \cmark
& Interactive-T  & All & Binary, List  \\
%\hline
\rowcolor{Gray} 
\erasmus\cite{erasmus}+\lisaalpha\cite{lisa}& 2018 &\pmf& % dependant on \lisaalpha = spanning-tree
% (\TopST)
% & \DynHDynamic &Centralised &
                              Verifier & Verifier % & \xmark  & %Noninteractive-C  
& Noninteractive& All & List \\ %DoS-V
%\hline 
\erasmusod\cite{erasmus(extended)}+\lisaalpha\cite{lisa}& 2018 &\pmf& % (\TopST)% dependant on \lisaalpha = spanning-tree
% & \DynHDynamic &Centralised & 
                              Verifier& Verifier % & \xmark 
& Interactive-T & All & List  \\ %DoS-V
%\hline
\rowcolor{Gray}
\consensus\cite{cons} & 2018 &\pmf&Prover$^*$ & Self-LC % & \xmark  & %Noninteractive-U  
& Noninteractive&All & List  \\ %DoS-V
%\hline 
\pads\cite{pads} & 2018 &\pmf&Prover$^*$ &Self-LC % & \xmark  & %Noninteractive-U 
& Noninteractive& All & List   \\ %DoS-V
%\hline 
\rowcolor{Gray}
\salad\cite{salad} & 2018 &\pmf& Verifier & Self-V % & \xmark 
& Interactive-T & Sample & List \\ %DoS-V
%\hline
\wise\cite{wise2018} & 2018 &\pmf & Verifier & Verifier % & \xmark
& Interactive-T  &Sample & List \\
%\hline
\rowcolor{Gray} 
\slimiot\cite{slimiot2018}& 2018 &\pmf& Verifier& Verifier % & \cmark
& Interactive-T  & All & List \\
%\hline 
  \usaid\cite{us-aid}& 2018 &\pmf&  \xmark & % Subnet 
  Neighbour
% & \cmark
& Interactive-U  & All & List \\
%\hline 
%\attauth is based on swatt so it is \pmr
\rowcolor{Gray}
\attauth\cite{attauth} & 2018 &\pmr&Verifier & Verifier % & \cmark
& Interactive-T & Sample & Binary % (random)
  % (random)
\\
%\hline 
\mvs-D2D\cite{mvs} & 2018 &\pmf& Prover$^{*}$ & Verifier % & \xmark 
& Interactive-U  & All & List  \\
\rowcolor{Gray}
\mvs-Consensus\cite{mvs} & 2018 &\pmf&  \xmark & Verifier % & \xmark
& Interactive-U  & All & List \\
\museda\cite{museda} & 2018 &\pmf & Verifier & Neighbour % & \xmark 
& Interactive-T  & All &List  \\
%\hline
\rowcolor{Gray}
\sap\cite{sap}& 2019  &\pmf, DM& Verifier& Verifier % & \xmark 
& Interactive-T  & Sample & Binary  \\ 
%\hline 
%DoS-v
\mtra\cite{mtra} & 2019 &EM  & Verifier &Self-L % & \xmark 
& Interactive-T & All & List \\ %Page 2 of ESDRA claims it is hw-based?
%\hline 
%\radis uses symmetric keys to mitigate DoS 
\rowcolor{Gray}
\radis\cite{radis} & 2019  &\pmf, DM, SF& Verifier & Verifier % & \xmark 
& Interactive-T  & All & Binary  \\
%\hline 
%I --> accusation report ? or it is List ? based on the accused ones, it considers others healthy.
\esdra\cite{esdra} & 2019  &\pmf& (2)  & % Reputation
                                                                              Neighbour  % & \xmark 
& Interactive-U  & % low-credit
Sample &List \\
\rowcolor{Gray} 
\eapa\cite{eapa} & 2019 &\pmf& Verifier & Neighbour % & \cmark  & %Noninteractive-U  
& Noninteractive& All& List  \\
% \hline
\shela\cite{shela} & 2019 &\pmf & Verifier$^{\dagger}$ & (3) % & \xmark
& Interactive-T & All & List \\
% \hline 
\rowcolor{Gray}
\healed\cite{healed} &2019 &\pmf& \xmark & Neighbour % & \xmark 
& Interactive-U  & Sample & List \\
% \hline
%am not sure about DoS in SADAN
\sadan\cite{sadan}+\diat\cite{diat} & 2019 &DM, CF& \xmark & Jury % & \xmark 
& Interactive-U  & Sample & List  \\
% \hline
%in SFS, the third party (cloud server) who does the main computations is untrusted, but Verifier is trusted
\rowcolor{Gray}
\sfs\cite{sfs} % \ \ (*)
                 & 2019 &\pmf& Verifier & Self-V % & \cmark
& Interactive-T  & All &List  \\
% \hline
\pasta\cite{pasta} &2019 &\pmf& \xmark & % (2) 
                                         Neighbour % & \cmark
& Interactive-U  & All & List  \\
% \hline 
\rowcolor{Gray}
\dads\cite{dads} & 2019  &\pmf&Verifier & Neighbour % & \xmark 
& Interactive-T  &All & List  \\
% \hline 
\simpleplus~\cite{simple} & 2020 &\pmf&Verifier & Self-V % & \xmark 
& Interactive-T  & All & List  \\
% \hline 
\rowcolor{Gray}
\cora\cite{cora}& 2020 &\pmf& Verifier& Verifier % & \xmark
& Interactive-T  &All & Binary, List  \\
% \hline
\sara\cite{sara} & 2020 &\pmf, SF& Verifier& Verifier % & \xmark
& Interactive-T  & Sample &List \\
\rowcolor{Gray}
\fadia\cite{fadia} % \jlp{I think DA should be ticked for this}
                 & 2021  & \pmf& Verifier& Self-L % & \cmark  & %Noninteractive-U  
& Noninteractive& All & Binary, Intermediate, List  
\\
% \hline
\bfb\cite{bfb} % \ \ (*)
                 & 2021 & \pmf& Verifier & Self-LC % & \xmark & %Noninteractive-U  
& Noninteractive& All & Binary \\
% \hline 
\rowcolor{Gray} 
\mton\cite{m2n} % \ \ (*)
                 & 2021 & \pmf& \xmark & Neighbour % & \xmark
& Interactive-U  & All & List \\
% \hline
\swarna\cite{swarna} & 2022 & \pmf & Verifier& Verifier % & \xmark 
& Interactive-T  & All  &List
\\
% \hline  
\rowcolor{Gray}
\shots\cite{shots} & 2022 & \pmr & Verifier& Verifier %& \xmark
& Interactive-T  & All & List 
\\
% \hline
\hola\cite{hola} & 2022 &\pmf & \xmark & 
                                         Neighbour % & \cmark
& Interactive-U  & All  & List  \\
% \hline
\rowcolor{Gray}
\fesa\cite{fesa} & 2022 & \pmf& Verifier$^{\dagger}$ & Verifier  % & \xmark  & %Noninteractive-U  
& Interactive-T& Sample & List \\
% \hline
\scraps\cite{scraps} & 2022 &\pmf, SF & \xmark & %blockchain/ proxy Verifier
% smart contract, 
                                                                         Proxy Verifier % & \xmark 
& Interactive-T  &Sample & List \\
\hline
\end{tabular}}
\end{center}
% \vspace{-18pt}
\end{table*}

% \raggedright 
%---: The metric is not discussed in the associated paper. % \quad \cmark: physically captured devices are detected. \quad \xmark: physically captured devices are not detected.
%\\
% (\TopST): It denotes a spanning tree that is dependent on the underlying CRA protocol. \\
\begin{note}[\cref{table:categories}] The following provides some
  additional notes on \cref{table:categories}.

  \begin{itemize}[leftmargin=15mm]
  \item [\xmark:] The protocol does not use a delegated trusted party.
  \item [Provers$^{*}$:] In \consensus and \pads, attestation includes a convergence phase where all provers obtain information about all other provers. The relying party can obtain the attestation result by querying {\em any} of the provers after convergence. In \mvs-D2D the delegated trusted party is the initial prover.
  \item [Verifier$^{\dagger}$:] These protocols additionally use edge verifiers. The edge verifiers are untrusted in \shela (and must be attested separately), but trusted in \fesa.
    
  \item 
    [(1):] The validator is the Verifier in \seed but Self-L in \seda, and it is not clear from the paper~\cite{seed} who the validator is when the protocols are coupled. 
  \item 
    [(2):] \esdra uses cluster heads chosen by a reputation mechanism to make decisions about provers. However, cluster heads' reputation may change over time. 
  \item 
    [(3):] The validator in \shela is the same as the validator of the underlying RA protocol. % \bd{
  \end{itemize}
  % Why is SAP both All and Sample?}\sa{because the attestation set is a subset (possibly equal) of all provers. It is not a random sample or only bad reputation ones.}
  % \bd{Is Hola \pmf or \pmr}\sa{when it does not say anything about random addresses, we can conclude that it means \pmf.}
\end{note}

\subsection{Attestation Target}\label{subsec:threats to prover integrity}

A prover device is typically specified as a single entity in CRA
protocols but, in most cases, it logically consists of two entities: a
\emph{trusted environment} (e.g., \emph{Trusted Platform Module}
(TPM)\cite{book-TPM}, \emph{Trusted Execution Environment}
(TEE)~\cite{tee}, \trustlite~\cite{trustlite}, \smart~\cite{smart}),
which serves as the \emph{root of trust}, and an \emph{untrusted
  environment} to be verified~\cite{hola}. 
% , since such a code is always untrusted.
We give a summary of these environments for the reviewed protocols in
the Appendix (\cref{app:architecture}).
Attestation code running inside the trusted environment is
assumed to %, and expected to work correctly (i.e.,
produce correct results except under the physical adversary
threat model (\cref{subsec:threats to CRA integrity}).
In the latter case, 
% one often introduces additional assumptions
weaker, conditional trust assumptions are made, e.g., that the secure environment does not act
arbitrarily or reveal secrets unless a physical adversary (see~\cref{subsubsec:adversary-models} for the definition of the adversary) takes the
prover offline~\cite{darpa}. We discuss detection and mitigation measures against physical adversaries in \cref{protection and mitigation}.

The target of attestation is generally the memory in the untrusted
environment. However, there is some variety on exactly what is
measured in CRA protocols.
%%% the following is a tautology:

%
%\sa{Pure software CRAs are often an
%exception as there is no hardware to create a trusted environment~\cite{swarna}.}

% \jlp{Justify this section's inclusion}
% CRA protocols may attest different parts of each prover.  
% Most CRA protocols concern the provers' \emph{program memory} and some
% concern {\em data memory} or the {\em entire memory}.
% %(e.g., \pads~\cite{pads}, \fadia~\cite{fadia}, \swarna~\cite{swarna}, \dads~\cite{dads})
% %\cite{simple,erasmus,pads,seda,fadia}
% %%%% SR: I find the following sentence controversial and too specific to be
% %%%% relevant here, I would prefer to remove it:
% %% Although attesting the entire memory provides a high level of integrity assurance,
% %% attesting only a set of (possibly random) memory locations offers
% %% improved protection when the adversary may know the healthy
% %% measurement for the entire memory in advance and is able to replace a
% %% measurement by the healthy one (e.g., \mtra~\cite{mtra}).
% Two other possible attestation targets are the {\em service-flow} or the {\em
%   control-flow} of an application program running on provers. 
% % (e.g., \radis~\cite{radis},
% % \sara~\cite{sara},\scraps~\cite{scraps}).
% In the following we give details about these attestation targets and
% classify CRA protocols according to them in the ``Attestation Target''
% column of~\cref{table:categories}. %  and
% \cref{app:attestation target} for more details).  which determines the
% attestation target of each CRA).
\begin{description}[itemsep=2pt,leftmargin=0pt]
\item [{\em Program-memory} (PM)] attestation generates a measurement
  for the entire region in which the prover's program (binary)
  instructions reside (e.g., \dads~\cite{dads}). Some protocols
  generate the measurement by reading through the program memory
  locations in a randomised order (e.g., \attauth~\cite{attauth} and \shots~\cite{shots}).

%   \bd{text needs improving} This way, it is not possible for an adversary to predict
% which memory location is accessed due to pseudorandom
% memory access. Therefore, the measurements of even two provers 
% that are exactly the same and are supposed to be running the same 
% software will be different and an adversary cannot 
% use the attestation result of a healthy prover for an unhealthy one  (e.g., \attauth~\cite{attauth}).

%   \bd{how do we determine
%     validity of measurements for \pmr?}
% \sa{The definition is wrong. It is not attesting .} \bd{Then can you fix the defintion? In this case, I suggest removing distinction between \pmf and \pmr }

\item [{\em Data-memory} (DM)] attestation aims to protect against
  compromise of the prover's data memory at run-time (e.g.,
  \sadan~\cite{sadan}). %  by % generating a measurement 
  % % of the
  % data memory
  % region.

\item [{\em Entire-memory} (EM)] attestation aims to protect against
  malicious code injection by measuring the entire memory
  including % the free memory space. Such protocols may attest
  % the entire memory (including 
  the data and program memory and free space. A protocol may first
  fill the free space with random values (unknown to the adversary)
  before taking a measurement. As with PM above, EM protocols may
  also % the entire memory set locations (EM-F), or
  attest a (possibly random) memory set of locations (e.g.,
  \mtra~\cite{mtra}).  % (EM-R). \bd{There's only one EM protocol and it's EM-R. Why bother with the distinction of EM-R?}

%\footnote{In this paper, we assume PM or EM is full unless they explicitly mention random memory addresses.}.
%Although attesting the entire memory
%offers a high level of integrity, attesting only a set of (possibly
%random) memory locations offers improved protection when the adversary may know the healthy measurement for the entire memory in advance and is able to replace a measurement by the healthy one (e.g., \mtra~\cite{mtra}).

\item [{\em Control-flow} (CF)] attestation aims to protect
% against  adversaries that can manipulate %data memory 
  the control-flow of instructions at run time, defending against
  tampering of the runtime stack or heap may be tampered with (e.g.,
  causing the order of instructions that are executed to be modified),
  without changing the binaries. There are many control-flow protocols
  for
 % RA~\cite{c-flat,lo-fat,tiny-cfa,litehax,hafix,diat,atrium,do-ra,oat}, 
  RA~\cite{c-flat,lo-fat,litehax,do-ra,oat}, 
however,
  to the best of our knowledge, the only CRA in the category is \sadan~\cite{sadan}, which is built on a data integrity RA, called \diat~\cite{diat}.

%In a swarm, each functionality offered by
%a specific prover is performed by an independent software
%component called {\em service.} A service operates  
%legitimately when it is not maliciously affected 
%directly or indirectly by the previous interactions among services. 
%A subset of services across a swarm may interact among themselves to compose a distributed service. The communication data
%exchanged among previous service interactions affect
%the current %state 
%operations of a prover.  Therefore, an attestation target could be
%the interactions and the data exchanged during these interactions to see if there is any non-intended operation due to these interactions with the infected
%device~\cite{sara}. Service flow attestation (defined below) capture this.

\item [{\em Service-flow} (SF)] attestation aims to protect against
  adversaries that
  % may let binary code remain unchanged, but they
  change the performed service, which is an expected functionality of a prover performed by an independent software. In this case, the attestation verifies the
  flow of data communicated between
  devices to assure that provers are not taking actions based on 
data from malicious services (e.g., \radis~\cite{radis}, \sara~\cite{sara}).

% \item [Topology (Top)] This is called Full-QoSA ... \sa{I want to remove this as only one protocol mentioned it.}

\end{description}

\subsection{Trustworthiness of Verification} \label{subsec:threats to verifier
  integrity}

Devices executing a CRA protocol may take on more than one role. For
example, a device may be both a prover and a verifier~\cite{cons},
both a prover and an aggregator~\cite{simple}, or both a verifier and
an aggregator~\cite{shela}. In many protocols, the role of the
aggregator is implicitly included in the prover's
role~\cite{simple}. This requires a careful consideration of the trust
assumptions that are made in the CRA protocol. 

Ultimately, unhealthy devices must be identifiable
to a \emph{relying party}~\cite{10.17487/RFC9334}. % (e.g., the
                                % network owner or operator).
Our trust assumptions are therefore made 
%Our definitions of trust are therefore described
from the perspective of this relying party.

The relying party may assume a special entity to provide a trustworthy
judgement on the healthiness of provers
(see~\cref{table:categories}). In this paper, we refer to this entity
as a {\em Delegated Trusted Party} (DTP). In RA protocols, the DTP is
usually referred to as a \emph{verifier}, which receives a measurement
from the attestors (provers) and provides a result to the relying
party~\cite{10.17487/RFC9334}. Following this idea, the most common
approach in CRA protocols is to designate a special {\it verifier} to
perform the attestation (e.g., \seda~\cite{seda},
\sara~\cite{sara}). However, CRA protocols that use centralised
verifiers may have a single point of failure, scale poorly or require
maintenance of a specific %spanning tree
network topology, %(see \cref{subsec:topology})
which may not be possible in
dynamic networks.  Therefore, a variety of other mechanisms for
building trustworthiness of verification have been developed.

A CRA verifier (when used) may store information about the swarm, such
as the number of provers, shared and private keys with the provers,
expected measurements, etc. As such, a CRA verifier is often assumed
to be a more powerful computational device than the provers, with a
greater range of capabilities~\cite{mtra}. Even when a verifier is
used, the functionality assumed of a verifier is not fixed, and often
differs from an RA verifier. For example, in SIMPLE+, a prover
generates its own attestation results, and the verifier is only tasked
with collecting these and reporting them to the relying
party~\cite{simple}.

We therefore consider the task of {\em validating} a measurement to be
separate from the verifier. As we shall see below (and in the {\em
  Validator} column of~\cref{table:categories}), judging the
acceptability of a measurement may be carried out using a range of
techniques, e.g., in \consensus~\cite{cons} and \pads~\cite{pads} all
provers share the attestation results of all other provers and
converge to a decision about the entire swarm. After convergence, the
relying party may query any of the provers to obtain the attestation
result. \shela~\cite{shela} and \fesa~\cite{fesa} use edge verifiers
and \esdra~\cite{esdra} uses cluster heads to split the swarm into
sub-networks. \fesa~\cite{fesa} assumes that the edge verifiers are
trusted, but \shela~\cite{shela} and \esdra~\cite{esdra} ensure
trustworthiness by periodically attesting the edge
verifiers~\cite{shela} or using a reputation score over the cluster
heads to determine whether they need to be attested~\cite{esdra}.

%   or provers whose trust has been
% established~\cite{us-aid}.

% A CRA protocol that employs 
% edge verifiers or cluster heads,
% either
% assumes that the edge verifiers (or cluster heads) are
% trusted~\cite{fesa} (in which case they do not need to be attested and part of DTU) or
% untrusted. When the edge verifiers (or cluster heads) are untrusted, the trustworthiness of verification is again based on the trusted verifier, as they
% are either attested by the trusted verifier~\cite{shela} or attested when they
% have a low enough \emph{reputation}~\cite{esdra}.

%One approach is for the attestation to be performed by multiple
%devices
%such as neighbouring provers~\cite{cons,pads}, edge
%verifiers~\cite{shela} and cluster heads~\cite{mtra}.

%In some CRAs with distributed topology (see \cref{subsec:topology}), 
%a wide range of
%trust-establishment techniques have been developed. Here,
%when neighbouring
%provers are used, one may typically 
%%set up a trusted verification \emph{subnetwork}~\cite{us-aid}, 
%perform attestation through
%\emph{consensus}~\cite{pads} or \emph{jury}~\cite{sadan}. 
% In some CRAS, attestation includes a convergence phase where all provers obtain information about all other provers. The relying party can obtain the attestation result by querying any of the provers after the convergence.

Several protocols (e.g.,~\cite{us-aid,mvs,healed,pasta}) do not use a
DTP, and instead assume that validation is carried out by a (group) of
other provers, e.g., using a consensus protocol to build up
trustworthiness of the result. In general, after a prover generates a
measurement there are many options for validating (i.e., verifying in
RA terms~\cite{10.17487/RFC9334}) whether it matches an expected one
(i.e., a healthy status) and producing a trusted result. We classify
these as follows.% \mbox{Self} (e.g., \pads~\cite{pads}) and \mbox{neighbour} protocols (e.g.,
% \pasta~\cite{pasta}) suit settings where provers have sufficient
% memory to store the expected measurements, and enough computational
% power to compare their attestation results, thus, leaving less work
% for a verifier to do. 
% %\sa{A neighbour prover can be 
% %physically neighbour (in the same communication range) such as in \usaid~\cite{us-aid} or a logical neighbour such as in \hola~\cite{hola},
% %where each device is identified by an m-bit number computed by a hash function; and using these identifiers, devices are linked to their predecessors and successors, thus creating a ring.}
% CRA protocols with no central
% verifier apply neighbour attestation to
% distribute the attestation responsibility to all provers (e.g., \hola~\cite{hola}). 
% \mbox{Self-V} protocols enable a verifier to send the updated expected measurements to provers as needed, but with an additional communication cost since such measurements are included in every attestation request.
% \mbox{Verifier} protocols are most widely used CRAs for
% low-end resource
% constrained devices~\cite{Ambrosin}, since a verifier with additional memory and computation power
% can be chosen (e.g., \cora~\cite{cora}). \mbox{Jury} protocols are recommended
% when there is no trusted verifier (e.g., \sadan~\cite{sadan}).
\begin{description}[itemsep=2pt,leftmargin=0pt]
\item [{\em Self}] protocols, where the measurement is validated by
  prover device itself. 
  % The prover's attestation result indicates the output of the
  % validation (typically Boolean flag indicating whether the prover is
  % healthy). 
  The prover therefore needs to know the  expected
  measurements, and here there are two options: (1) {\it Self
    verifier} (Self-V) protocols, where a verifier sends the set of
  expected measurements to provers, and (2) {\it Self local} (Self-L)
  protocols where the expected measurements are stored locally within
  the prover itself.  Both Self-L and Self-V protocols assume a trust
  anchor (see \cref{subsec:architecture}) within the provers
  themselves to ensure trustworthiness of the validation.  When such a
  root of trust is not available, a CRA protocol may run a consensus
  algorithm to come to agreement about on the validation
  results. We categorised these as {\it Self-Local Consensus}
  (Self-LC) CRA protocols.
\item [{\em External}] protocols, where the provers send their measurements
to a third party for validation. %that performs the validation.
We classify these
%into subcategories: 
as follows: (1) {\it Verifier} protocols, where the measurements are
validated by a centralised verifier, (2) {\it Neighbour} protocols,
where the measurements are validated by the prover's neighbouring
provers, and (3) {\it Jury} protocols, where the measurements are
validated by a jury of provers. The recently developed
SCRAPS~\cite{scraps} protocol delegates validation to an untrusted
{\em proxy verifier} implemented as a smart contract on a
blockchain. The proxy verifier interacts with the (trusted)
manufacturer smart contract to perform validation.
% (4) {\it Concensus (C))} protocols, where the validaotors need to use concensus to come up with a single  
\end{description}

\input{threat_models.tex}

\section{Protection and Mitigation}\label{protection and mitigation}

In this section, we
identify how CRA protocols mitigate software threats using their
underlying minimal secure hardware (see~\cref{subsec:architecture}) and consider scenarios where the
secure hardware itself is physically attacked, and present
% via examples
a systematisation of techniques found in
the literature to mitigate or detect such attacks (see~\cref{subsec:mitigation-detection}).
%, designed and used for such attacks
Finally we discuss DoS attack mitigation (see~\cref{subsec:DoS mitigation}).

\subsection{Defending Against Software Attacks}\label{subsec:architecture}

As discussed in~\cref{subsec:threats to prover integrity}, CRA
protocols assume that provers are also equipped with a {\em root of
  trust}~\cite{healed,us-aid,hola,scraps} to protect against
software~attacks.  CRA protocols are designed in three different ways, based on
assumptions on the underlying architecture of the root of trust of the
provers: software, hardware, or hybrid. Software-based protocols
(e.g., \cite{kovah,viper,sake,scuba,swatt,pioneer}) do not have any
specific hardware requirements; thus they are cheap, however, they
rely on strong assumptions that may be hard to achieve. These
assumptions themselves are the root of trust for software protocols.
Hardware-based protocols
(e.g., %~\cite{pufatt,Schellekens,schulz,flicker,copilot,bootstrapping,lightweight}
\cite{pufatt,lightweight,flicker}) rely on a secure hardware such as
TPM~\cite{book-TPM} or \emph{Physically Unclonable Functions}
(PUFs)~\cite{PUF} where attestation code is executed in a secure
environment.
%They are suitable for advanced computing platforms, e.g., smartphones or laptops, since
Such hardware components play the role of the root of trust but may be
too complex and expensive for medium- and low-end % embedded
devices.

Almost all CRA protocols are implemented on hybrid RA architectures,
e.g., SANA~\cite{sana} is implemented on SMART~\cite{smart} and
TyTAN~\cite{tytan}, EAPA~\cite{eapa} on TrustLite~\cite{trustlite},
and FeSA~\cite{fesa} on HYDRA~\cite{hydra}. Hybrid architectures
combine software and a minimal amount of hardware to create a
root-of-trust~\cite{minimalist}. This hardware may include a
\emph{Read-Only Memory} (ROM) to prevent modification of attestation
code and a \emph{Memory Protection Unit} (MPU)~\cite{MPU} to guarantee
non-interruptability (aka atomicity) of the attestation procedure % ,
% controlled invocation of the procedure (to ensure the protocol starts
% from the beginning), and 
as well as privileged access to secret keys, session ids, nonces,
counters, time stamps, and clocks. For example:
% \darpa~\cite{darpa} and 
\begin{itemize}[leftmargin=8pt]
\item \seed~\cite{seed} assumes hardware-protected {\it Reliable
    Read-Only Clocks} (RROCs) that are (loosely) synchronised within
  devices running the CRA protocol.  Without an RROC, \advsw may
  manipulate the clock and pre-generate healthy attestation results
  for future use when the prover is actually unhealthy.
\item \erasmus{}+\lisaalpha~\cite{erasmus,lisa} and \seed~\cite{seed} assume
  hardware-protected {\em Pseudo Random Number Generators} (PRNGs)
  that are initialised with a secret seed and used to generate
  attestation times. Without these, an \advmsw that knows the time
  interval between attestation rounds may be able to hide by deleting
  itself prior to measurements being taken. The PRNG and secret seed
  are used to perform attestation at random time intervals that cannot
  be predicted by \advmsw.
\end{itemize}
% To prevent this, a protocol may , yet are known in advance to
% the devices running the protocol. This may be achieved via Both the
% PRNG and the seed must be protected by the trusted environment of the
% devices.

% prevents 

% \seed~\cite{seed}
% on devices, which is stored in ROM and 

Ammar et al.~\cite{simple}, propose a design that deviates from the
above classification. They implement a root of trust using a formally
verified software-based memory isolation hypervisor ($S\mu V$) to
implement a root of trust that claims to provide the same guarantees
as traditional architectures, without specific hardware requirements.

\input{physical_attacks.tex}

%% file: threat_models.tex
\subsection{Threat Models} \label{subsec:threats to CRA integrity} An
adversary can interfere with inter-device communication and compromise
any untrusted ones -- being able to read their memory and modify their
behaviour. In this section, we discuss possible attacks on CRA
protocols, then present adversary models to capture different attacker
capabilities.

\subsubsection{Software Attacks} \label{subsubsec:software-attacks}
By exploiting software vulnerabilities, a remote attacker can read from or write to any unprotected memory regions on a device. As well as enabling the attacker to execute arbitrary code, the following are examples of attacks that may be performed on an attestation protocol if there are no hardware (such as \smart~\cite{smart} or \trustlite~\cite{trustlite}) or software restrictions (such as in \simpleplus~\cite{simple}) in place to prevent it (i.e., if the attestation code is in the untrusted rather than trusted environment,~\cref{subsec:threats to prover integrity}).

\begin{itemize}[itemsep=0pt,leftmargin=8pt]
    \item {\it Illegitimate results}, where an attacker may modify the measurement algorithm, which may result in the prover generating an incorrect attestation result~\cite{smart, simple}.
    \item {\it Forged results}, where an attacker may read the secret keys of a prover and forge a measurement for that \mbox{prover~\cite{lisa,smart}}.
    \item {\it Pre-generated results}, where an attacker may trick a prover into generating measurements of the current (healthy) software state for future use by manipulating the prover's clock. These measurements are then used as responses during an attack, when the true software state is unhealthy~\cite{seed}. 
\end{itemize}

\subsubsection{Physical Attacks} \label{subsubsec:physical-attacks}
If a device is physically attacked, the attacker can modify attestation code or extract information such as secret keys, even if restrictions exist to protect these from software-based attacks. A physical attack can either be intrusive or non-intrusive.
% \begin{itemize}[itemsep=0pt]
% \item 
\emph{Non-intrusive attacks} include fault injection or side-channel
attacks to extract a device's secret keys without interfering with its
normal operation~\cite{tamper,abera,physical}. This generally requires
the help of a physical tool (e.g., an oscilloscope or
EM-receiver/demodulator) that is attached to the victim prover or
installed in close proximity~\cite{darpa}. \emph{Intrusive attacks}
involve taking the target device offline for a non-negligible amount
of time to extract information and/or modify a device's software
and/or hardware.
% \end{itemize}

  \subsubsection{Network Attacks} \label{subsubsec:network-attacks}
  Without compromising any provers, an attacker can exploit any
  vulnerabilities in a CRA protocol at the network level. For example,
  as described by Ibrahim et al.~\cite{seed} this includes the
  following attacks.
\begin{itemize}[itemsep=0pt,leftmargin=8pt]
    \item {\it Delay attacks}, where an attacker delays an attestation
      request to evade detection until the attacker's goal is achieved.
    \item {\it Dropping attacks}, where an attacker drops every
      report that does not indicate a healthy status to prevent
      detection of compromised provers. %, so the verifier
%      cannot detect compromised provers when receiving a
%      report.
    \item {\it Record and replay attacks}, where an attacker 
      records and drops healthy attestation results and later replays
      them % during an attack
      to make %deceive the verifier into considering
      an unhealthy prover appear healthy.
\end{itemize}
To avoid such attacks, a protocol can utilise nonces, counters, or
timestamps stored in a secure memory~\cite{brasser}.
In the case of timestamps, provers and verifiers must have synchronised clocks to provide them with attestation times. 

\subsubsection{Adversary Models} \label{subsubsec:adversary-models}

CRA adversary models have been inherited from RA adversary models,
e.g., Abera et al.~\cite{Abera-Invited}, and 
further discussed
%previously in the literature
in the CRA context in~\cite{Ambrosin, seed, brasser}. % We present
The following adversary capabilities summarise the capabilities found
in the CRA protocol literature and 
are written in the context of the trusted/untrusted environments described
in~\cref{subsec:threats to prover integrity}. 
The mapping of the reviewed CRA protocols to these
adversary models is shown in~\cref{table:adversary models}. 

% \jlp{Do you think this section needs a bit more explanation surrounding the models? I think it might benefit from it but on the other hand it might be a waste of space.}

\begin{description}[itemsep=2pt,leftmargin=0pt]
\item [{\em Software Adversary}] (\advsw), which can modify any memory
  in the untrusted environment. However, once it has done so, this
  adversary cannot restore the memory to the original (presumably
  expected) state.
\item[{\em Mobile Software Adversary}] (\advmsw), which inherits the
  abilities of \advsw, but \emph{can} restore memory to its original
  state.  \advmsw may use its abilities to try to avoid
  detection. E.g., if a piece of malware knows when a prover is going
  to perform attestation, it can erase itelf to make the device appear
  healthy when attestation is performed.
\item [{\em Physical Non-Intrusive Adversary}] (\advpni), which can
  read from but not write to \emph{any} memory location on the
  target device, including memory in the trusted environment. 
\item [{\em Physical Intrusive Adversary}] (\advpi), which can read
  from and write to \emph{any} memory location on the targeted
  device, including memory in the trusted environment by taking a
  device offline for some amount of time.

% \item [\em Physical Non-Intrustive ]
%   \sa{The CRA protocols only consider physical adversaries that capture provers for a non-negligible amount of time.

%     The protocols may use different techniques (e.g., applying a heartbeat protocol) to detect offline provers. But, the CRA protocols assume that physical non-intrusive adversaries are out of their scope, since on their own and

%     without . Following the literature, we assume that such adversaries are out of the scope of this paper.}

\item [{\em Dolev-Yao Adversary~\cite{dolev-yao}}] (\advc), which is assumed to have full control
  over the communication network. It can 
  eavesdrop on, modify, delay, drop and inject messages. %  in the network.

\end{description}

%\jlp{Ambrosin et al. also have an $Adv_{SPI}$ that we do not have, and I am not sure why we do not have it.}

%%% Local Variables:
%%% mode: latex
%%% TeX-master: "main"
%%% End:

%% file: physical_attacks.tex
\subsection{Defending Against Physical Attacks}\label{subsec:mitigation-detection}

A physical attack %on a device in a CRA protocol 
allows an attacker to bypass %all of 
a device's hardware %protections.  
% \jlp{Seems a bit unfair given PUFs and physical defences exist}
%Among such an adversary's capabilities is the capacity to bypass hardware
restrictions to gain access to any secrets stored on a device. These secrets can  %either 
be used to impersonate the captured device, or malware on the device
can perform arbitrary operations using the secrets.

%
%There are three ways that a CRA protocol can handle physical attacks: (1) not at all, (2) with mitigation, (3) with detection.

%\subsubsection{Ignoring Physical Attacks}
A CRA protocol %that provides zero 
with no protection against physical %intrusive 
attacks may leave the entire swarm vulnerable if just one prover is physically compromised. 
An example is \simpleplus~\cite{simple} where two symmetric %authentication and attestation 
keys are shared by all provers and a verifier. 
The physical compromise of any prover gives the adversary access to both keys, with which it can impersonate the entire swarm and send the verifier seemingly legitimate %attestation 
reports as %there is no way for 
the verifier %to 
cannot tell whether this has occurred.  

% In swarms with heterogeneous inter-connected devices, some devices may
% be physically protected, some others may not. A physical attack even
% to a single device may affect the attestation results of the
% others.

% CRAs predominantly consider non-intrusive physical adversaries
% to be out of their scope\footnote{\salad~\cite{salad} is an exception
%   which considers side-channel attacks, but it does not explain
%   whether it is secure against them.}, 
% \sa{since they are too strong and also require physical tools to be attached to the victim device, thus they are considered very likely to be detected.  
% However, some CRAs provide security guarantees against intrusive
% physical adversaries (e.g., \scapi~\cite{scapi}). Such CRA schemes either mitigate the attacks or aim to detect them.

%\SR{DOUBLE CHECK. According to our adversary capabilities, a
%  PI adversary can do what a PNI adversary does and more. So saying
%  ``predominantly consider non-intrusive adversaries out of scope''
%  does not match with the table, because the table shows many
%  checkmarks in the PI category which includes the PNI
%  capability. Calling \salad an exception for considering side-channel
%attacks indicates a mismatch between our adversary model and what
%protocols defending against PI try to achieve?  I suggest the
%following edits, but Sharar/Jay DOUBLE CHECK that they make sense.}
%\sout{CRA protocols predominantly consider \emph{non-intrusive} physical
%  adversaries to be out of scope}
Few CRA protocols consider a \emph{non-intrusive} physical
  adversary rather than a \emph{intrusive} physical adversary.\footnote{\salad~\cite{salad} is an
  exception, which considers side-channel
  attacks.} %. So, we assume this attack to
% be out of the scope of our paper and exclude it from
% ~\cref{table:security properties}.}, since they are too strong.
Although devices spread over a large area
are difficult to monitor, non-intrusive attacks require physical tools to be
attached to the victim device, making them difficult to
perform. % thus they are considered very likely to
% be detected.
% since devices deployed over a large area are difficult to moin
% monitored, detecting a physical attack would be
% difficult.
% \sa{since they are too strong, but 
% if a swarm of devices is deployed in a dynamic, wireless environment,
% it seems reasonable that an adversary might be able to tamper with a
% few of them. For example, environmental sensors spread over a large
% area that cannot be otherwise monitored (hence the need for this swarm
% of devices in the first place).  Moreover,   
%\sout{However, some CRA}
Several protocols do provide mitigation and detection
against \emph{intrusive} physical adversaries (e.g.,
\scapi~\cite{scapi}). % \footnote{\sa{Investigating hardware tamper
    % resistance~\cite{johansson2020tamper, akter2023survey}, industry
    % certification practices (e.g., Common
    % Criteria~\cite{CommonCriteri}), and open source designs (e.g.,
    % OpenTitan~\cite{OpenTitan}) are outside of the scope of the
    % paper.}}
% handle such attacks: (1) not at all, (2)
%\subsection{Physical Attacks}
%%CRAs that aim to protect against physical attacks either mitigate
%%them or detect them. 
% Such CRA schemes either mitigate the attacks or aim to detect them. 
%
%This leads to another metric defined as follows, based on which CRAs can be categorised (see the last column in \cref{table:categories}). 
%
We discuss existing mitigation and detection approaches next and show the mapping of
the reviewed CRA protocols to their protection strategies in the
middle column of 
\cref{table:security properties}. 
It is worth noting that in dynamic wireless environments with a
deployed swarm of devices, e.g., environmental sensors in remote unmonitored areas, an adversary could tamper with some
devices. %necessitating the swarm deployment.
However, the details of
hardware tamper resistance~\cite{johansson2020tamper, akter2023survey}, industry certification practices (e.g., Common Criteria~\cite{CommonCriteri}), and open source designs (e.g., OpenTitan~\cite{OpenTitan}) are outside of the scope of this paper.

% handle such attacks: (1) not at all, (2)
%\subsection{Physical Attacks}
%%CRAs that aim to protect against physical attacks either mitigate
%%them or detect them. 

\subsubsection{Physical Attack Mitigation}
The minimal defence against physical intrusive attacks is to prevent the compromise of one device from affecting the security of other devices. For example, \sana~\cite{sana} gives each prover its own asymmetric key to sign its report with. If a prover is physically compromised, giving its key to an adversary, the adversary can impersonate the compromised prover, but not the provers it has not physically compromised because they use different secrets. Mitigation techniques make it
difficult for adversaries to forge reports for compromised devices,
enhance security against invasive physical attack on conventional
memories, or prevent attackers from decrypting previous
messages.

% \jlp{I need to look at \attauth and \sfs but, at a glance, it seems these might all be same thing -- asymmetric cryptography}
\begin{description}[itemsep=2pt,
  leftmargin=0pt]
\item [{\em Mitigation via unique keys} (\uk).]  \salad~\cite{salad}
  authenticates each device's attestation report with a unique
  key. This protocol mitigates physical attacks by executing code
  inside devices' TEEs and authenticating reports with unique
  attestation keys, making it difficult for adversaries to forge
  reports of compromised devices.

  \item [{\em Mitigation via PUFs} (\puf).]
  Physically Unclonable Functions (PUFs) are functions embedded into circuits, producing different outputs in each device. They can be used to generate secrets, eliminating the need to store secrets in memory. The authors of SHOTS~\cite{shots} and \attauth~\cite{attauth} claim this makes their schemes resistant to physical attacks; however, their physical adversary can only read from and not write to memory, unlike ours (in~\cref{subsubsec:adversary-models}).
  As each PUF is unique, this mitigation inherits the properties of \uk.
  % \attauth~\cite{attauth} employs PUFs to generate on-chip secrets, eliminating the need for stored secrets and enhancing security against invasive physical attacks on conventional memories \jlp{Why do PUFs enhance security against invasive physical attacks?}. PUFs are noisy functions embedded in circuits, producing output dependent on input and device structure. Multiple queries may yield slightly different results due to environmental factors. 

  % \item [{\em Mitigation via forward secure signature scheme} (\fss).]
  % \jlp{To discuss on Tuesday. I believe this can be filed under \uk. The unique aspect of \fss is that previous reports remain secret. However, we do not discuss secrecy in this SoK. }
  % \sfs~\cite{sfs} allows public verification of aggregated reports without revealing the provers' identity. In case a device's private key being compromised, previous signatures/messages remain secure. \sfs uses a forward-secure signature scheme and puncturable encryption to ensure the authenticity of pre-exposure signatures. To support private key updates for  aggregators, \sfs combines puncturable encryption with hierarchical identity-based encryption to prevent attackers from decrypting previous messages. This enables computations to be outsourced to cloud servers while keeping sensitive information secure.
\end{description}

\subsubsection{Physical Attack Detection}\label{subsubsec:physical-detection}
% Notation:
% Scapi uses \delta_{attack}
% FADIA uses \delta_h (Where h stands for hardware and maps to their adversary A_h)
% slimIoT uses T_{adv}
% HolA uses T_a
% DARPA uses T_{cap}
% EAPA uses T_{cap}
% US-AID uses t_{phy}
% PASTA uses \delta_a
% Conclusion: There is no consistency, except they always start with either T or \delta.
% We follow convention by making up our own notation (Delta representing a time interval and PI reflecting the type of adversary we are dealing with)
\label{subsec:mitigation-detection-da} %%% keep this label with the device availability text below
A stronger defence against physical intrusive attacks is to detect
them. 
% \begin{description}[itemsep=0pt,leftmargin=0pt]
%     \item [Device availability (DA)] protocols can detect a physical attack where an adversary captures a device for a non-negligible amount of time, making it unavailable to perform attestation. 
% \end{description}
CRA protocols use different techniques to detect physical
attacks under the assumption that performing such an attack requires
the victim device to be taken offline for a non-negligible amount of
time~\cite{scapi,us-aid,eapa,darpa,hola,slimiot2018,fadia}, which we
refer to in this paper using \attacktime. %  \bd{give citations for other
%   notions} \jlp{I do not follow. None of the other papers discussing
%   \advpi seem to spend time justifying their notation. Does it matter?
%   (See LaTeX comments for list of notation used by other papers.)
% }. 
These techniques broadly fall in to three categories: frequent
heartbeats; frequent secret updates; and frequent
attestation\footnote{By frequently, we mean at least every
  \attacktime.}.

\begin{description}[itemsep=2pt, leftmargin=0pt]
  % DARPA: A distributed protocol where provers send heartbeat messages to one another, that are rebroadcast so all provers have their own log of heartbeat messages ideally from all other provers. I'm not sure how the verifier querying logs works with provers needing to give honest logs but we're agreed that this is \hb anyway so I won't put more time into it.
  % EAPA: A weird one in that it requires an attestation (not swarm attestation) protocol, but is itself just a heartbeat protocol. Provers send heartbeats to their neighbours and accuse missing neighbours. Accusers are attested using an RA protocol before their accusation is accepted by the operator. 
  % US-AID: "Each device sends a heartbeat to all neighbours demonstrating its presence. Each device records the list of devices from which it received a heartbeat and drops secure communication with all devices that did not send a heartbeat". Also has an attest part of the protocol, where devices randomly attest their neighbour's software state.
  \item [{\em Heartbeat protocol} (\hb).] A heartbeat protocol
    frequently checks that provers are online. If a prover is not reached for more than \attacktime, that prover is considered physically compromised. A heartbeat protocol may be standalone (e.g., \darpa~\cite{darpa}) or it may be incorporated into a CRA protocol (e.g., \hola~\cite{hola}, \usaid~\cite{us-aid}).
  \item [{\em Secret updates} (\su).] 
  % SCAPI: SCAPI frequently generates a new session key. If a device is physically attacked, it will miss at least one session key update. As the most recent key is required for participation in both the key update and attest phases of the protocol, a physically attacked device cannot be successfully attested. 
  A CRA protocol can frequently update the secrets required to participate in attestation (e.g., keys in \scapi~\cite{scapi}). If a prover fails to receive the updated secret (e.g., due to being offline), it can no longer participate in the protocol and be successfully attested. 
  \item [{\em Attestation} (\att).]
  % PASTA: If a prover fails to participate in token generation (i.e., attestation) for more than \attacktime, it is considered physically compromised.
  % FADIA: Attestation frequency is \attacktime / 2 so "two consequent attestations are always received within no longer than" \attacktime time. 
  If \attacktime is large enough, and the attestation time short enough, it is feasible to simply ensure that provers attest themselves more frequently than \attacktime. If a prover fails to attest itself, it may be presumed to be physically compromised. 
\end{description}

% HolA:
\hola~\cite{hola} uses a combination of \hb and \att, and assumes that the attestation protocol is executed significantly more frequently than \attacktime. If a device fails to participate in the attestation protocol, a heartbeat protocol begins until the device either responds, or is missing for $\geq \attacktime$. 
% slimIoT:
\slimiot~\cite{slimiot2018} uses a combination of \att and \su. Attestation is performed frequently but the attestation procedure also updates a secret. This both detects the absent device by attestation and prevents the captured device from participating in future rounds of the protocol.

\subsection{Defending Against DoS Attacks}\label{subsec:DoS mitigation}
Denial of Service (DoS) is a threat to all networked devices. It is
not within the capability of a CRA protocol to prevent all DoS
attacks, however CRA protocols should (and do) consider application
layer DoS attack vectors introduced by the CRA protocol itself (e.g.,
\lisa~\cite{lisa}, \seed\cite{seed}). Because provers in CRA protocols
are envisioned as low-power devices, performing attestation is an
expensive (and blocking) task for provers. If an adversary can induce
provers to attest more often than the protocol intends, they can
easily perform a DoS attack on the swarm, preventing devices from
performing their intended function. In the next section, we will
explore two ways of preventing such an attack, namely, designation of
\emph{Non-interactive} CRA protocols and the \emph{Initiator
  Authentication} security property.

%% file: IA.tex
\section{Interactivity and Initiator Authentication}\label{subsec:verifier authentication}

We now turn to \emph{initiator authentication}, the first security property described in
\cref{fig:roles}. It ensures that provers only perform attestation
if legitimate initiator(s) kick-start the attestation process.  As
provers are assumed to be low-power devices, attestation is
an expensive (and blocking) task for them and generally cannot be
interleaved with other tasks. An adversary that induces provers to attest more
often than planned or attest at inopportune times can disrupt the
services these devices are expected to provide.
Whether initiator authentication is an appropriate mechanism to
prevent such an attack depends on the CRA protocol's design. 
We therefore first discuss the notion of \emph{interactivity}
(\cref{subsec:interactive}) before defining initiator authentication (\cref{sec:init-auth}).

\subsection{Interactivity}\label{subsec:interactive}

Attestation procedures may be initiated in several ways.  The
categories below are inspired by Ambrosin \etal~\cite{Ambrosin}, who
distinguish between interactive and non-interactive CRA
protocols. Unlike Ambrosin \etal~\cite{Ambrosin}, who only consider
protocols with a {\em trusted} initiator, we additionally
consider protocols that may be initiated by an {\em untrusted}
device % , and hence
% distinguish between
% e
% provide a more fine-grained categorisation of interactive CRA
% protocols (see \cref{table:categories}) 
to better understand how
protocols ensure the legitimacy of attestation requests (summarised in
\cref{table:categories}).
% to see whether
% an attestation request is an eligible one, i.e., if it is actually
% needed to start an attestation procedure.
%, when the initiator is an (untrusted) device.
% This is mostly important when the initiator is an untrusted device.

\begin{description}[itemsep=2pt,leftmargin=0pt]
\item[\normalfont\textrm{In}] {\em Interactive} (aka
    on-demand) protocols provers
  perform attestation after receiving a specific attestation request
  message. We distinguish two subtypes: (i) {\em interactive
    trusted}, denoted \textbf{Interactive-T}, where requests originate from
  trusted devices and (ii) {\em interactive untrusted}, denoted
  \textbf{Interactive-U}, where requests may originate from {\em untrusted}
  devices.
\item[\normalfont\textrm{In}] \textbf{\em non-interactive} protocols provers decide for
  themselves when to perform attestation, e.g., by using a secure
  clock to trigger attestation at set points in time.
\end{description}
\smallskip

\noindent  
In Interactive-T protocols, attestation requests are by definition
assumed to be legitimate. However, in Interactive-U protocols, the
initiator may be a compromised neighbouring prover device.
A mechanism to ensure in this case that the attestation is only initiated when
desired, is to protect the initiation code within the initiating
prover's trusted environment %to ensure its validity
(e.g., \usaid~\cite{us-aid}).
Interactive-U protocols also use other techniques to counter
illegitimate attestation requests. \esdra~\cite{esdra} and
\mton~\cite{m2n} use a reputation mechanism to choose cluster heads
and only provers with a high enough reputation score may initiate the
attestation.  In \pasta~\cite{pasta}, the initiator sends attestation
requests after a timeout, but only after verifying that its software
state is trustworthy. Moreover, attestation requests are only sent to
prover devices that were known to be healthy in the previous
attestation round. In \sadan~\cite{sadan}, a device may \emph{blame}
other devices and request a jury to attest them. However, any prover that blames a healthy device will itself
be blamed by the jury of the current round and hence attested in the
next round.

% where if an unhealthy prover
% blames a healthy prover multiple times to use the blaming mechanism to
% overload the system with attestation requests, the protocol leads an
% unsuccessful blame to an automatic blame of the blamer prover by the
% jury of the current round.

% the provers are first arranged into a spanning tree with an untrusted prover selected as the initiator. 

% attestation is controlled by a
% token generation mechanism, based on a timeout. 

% % each device is set up once by
% % the relying party.  Afterwards, provers continually execute the token
% % generation phase, in which they repeatedly generate tokens.  
% % Each
% % token attests the software and hardware integrity of all provers that
% % participated in its generation, at the generation time. Simultaneous
% % to the token generation phase, all devices perform the token exchange
% % phase where devices distribute, verify, and validate the generated
% % tokens from the provers.
% Devices eventually use their verified and
% validated tokens to determine the integrity of all provers in the
% swarm.  In this protocol, the initial prover sends an attestation
% request only if it notices that the newest token in whose generation
% it participated is older than a specific token generation time
% interval.

Interactive CRA protocols may be used in settings where initiators
make decisions based on some event, e.g., when an intrusion detection
system shows that a prover is suspected of being
compromised~\cite{leveraging,trust}, or when a monitoring mechanism
requests attestation in response to demand
\cite{erasmus(extended)}. % , the
% initiator starts the attestation procedure.
Non-interactive protocols reduce the communication overhead of attestation requests (and hence energy costs and run-time).  
% The non-interactive CRA protocols may reduce energy cost, , and
% run-time \bd{How?}.
They also mitigate computational DoS attacks, as there is no
attestation request that originates from the initiator (e.g.,
\seed{}+\seda~\cite{seda,seed}). The frequency of an initiator’s
requests can also be
decoupled from the frequency of a prover’s measurements allowing for the implementation of a simple collection phase on the
provers (e.g., \erasmus~\cite{erasmus}).

% They make both provers and
% initiators rely on a {\em secure random seed} shared among all devices
% and protected by their root of trust. This seed is used to generate a
% pseudorandom sequence of attestation times using a secured RROCs
% (see~\cref{subsec:architecture}), preventing adversaries from
% anticipating the next attestation procedure~\cite{seed}, thus
% mitigating mobile software attacks.

% \bd{Again I don't understand the purpose of this next paragraph. Why is it important?}
% The non-interactive protocols are good fit in settings where the topology may change unpredictably during an attestation procedure (see~\cref{app:dynamicity} for more details on topology) and for devices working under real-time constraints,
% since the provers will not use their computational resources 
% to authenticate new requests and preparing %(aggregated)
% reports regularly instead of performing their main tasks. 
% % They may also decouple 

%%%%% SR: Removed because HT, SR do not understand it... 
% Note that the security properties we present in \cref{sec:init-auth} only apply
% to the provers that are selected to {\em perform} the attestation, and
% do not provide any guarantees for provers that are not attested. Given
% the range of strategies used to initiate attestation, such general
% guarantees may be difficult to state, let alone prove.

\subsection{Initiator Authentication}
\label{sec:init-auth}

% are between the prover(s) selected for attestation and the relying
% party.

To prevent an adversary from exploiting the initiation mechanism, it
is necessary that provers only perform attestation in response to
legitimate requests. We define \emph{initiator authentication} based
on Lowe's injective agreement~\cite{lowe} to ensure a prover only
accepts requests genuinely from the initiator (as defined
in~\cref{sec:overview}), and each unique request no more than once.  % \bd{per attestation round?}.

\begin{definition}[Initiator Authentication (\va)]\label{def:verifier authentication} 
  A CRA protocol guarantees \va to a prover $p$ iff, whenever $p$ completes a run of the protocol, apparently having received an attestation request from an initiator $i$, then $i$ has previously been running the protocol, sending an attestation request to $p$. Furthermore, each run of the protocol by $p$ corresponds to a unique attestation request sent by $i$. 
\end{definition}

\noindent On its own, \va only prevents replay attacks. To prevent the
above DoS attack, where the adversary replays attestation requests, the initiator $i$
must be honest. If $i$ is dishonest, it can spam a prover with
legitimate attestation requests, causing a DoS attack without
violating \va. In practice, this assumption may take the form of
requests coming from a trusted third party (e.g., such as in~\seda~\cite{seda}, \simpleplus~\cite{simple}, \sana~\cite{sana}), \lisa~\cite{lisa}), 
or some secure attestation trigger on the attested
device (e.g., such as in~\seed\cite{seed}, \pads~\cite{pads}).

To prevent legitimate attestation requests from being replayed, a CRA
protocol may employ standard techniques such as counters
(e.g.,~\salad~\cite{salad}), timestamps (e.g.,~\sana~\cite{sana}), or
nonces (e.g.,~\scraps~\cite{scraps}).

% \bd{Does this belong here?}  
Non-interactive protocols assume the existence of secure hardware on
the device to initiate attestation, such as \seed's timeout circuit
and non-maskable interrupt~\cite{seed}. Abstractly, in our
model~(\cref{fig:roles}), this device could be considered both an
Initiator and a Prover, but there is no actual request message
sent. As there is no message to forge, we assume \va cannot be
violated in non-interactive protocols. Ensuring the security of the
attestation trigger (abstractly, the honesty of the initiator) is a
hardware/software level problem, not a protocol level one.

%% file: security_properties.tex
\section{Attestation Correctness}\label{sec:security properties}
\newcommand{\ccmark}{\raisebox{0em}{\color{blue!60}\ding{51}}}

\newcommand{\advswcm}{\ccmark}
\newcommand{\advpicm}{\ccmark}
\newcommand{\advccm}{\ccmark}
\newcommand{\advmswcm}{\ccmark}

\begin{table*}[!t]
\footnotesize
\begin{center} 
  \centering\caption{\label{table:security
      properties}\label{table:adversary models} Summary of adversary
    models, physical protection and aimed security properties}
\resizebox{0.99\linewidth}{!}{
\begin{tabular}{|l|c|c|c|c|c|c|c|lll|} 
\hline 
\rowcolor{DarkGray!70}
  \bf Protocol Name & \multicolumn{5}{c|}{{\bf Adversary Model} (\cref{subsec:threats to CRA integrity})}   & \bf Physical  & \va & \bf Individual/Group& \bf Sync/ASync& \bf Weak/Strong    \\
  \cline{2-5} % \hhline{~|*4{-}|} 
  \rowcolor{DarkGray!70}
              & \advsw & \advmsw & \advpni & \advpi & \advc &  {\bf Protection}~(\cref{protection and mitigation}) & (\cref{subsec:verifier authentication})& {\bf Attestation} (\cref{sec:security properties}) & {\bf Attestation} (\cref{sec:security properties}) & {\bf Attestation} (\cref{sec:security properties})      \\
\hline 
\seda~\cite{seda} &  \advswcm & &  & & \advccm& \xmark&\cmark&  Group & Async& Weak   \\ 
\rowcolor{Gray}
  \darpa~\cite{darpa}+\seda~\cite{seda}  &  \advswcm & &  &  \advpicm & \advccm & \hb &\cmark & Group & Async & Weak  \\ 
\sana~\cite{sana}& \advswcm & &   &  \advpicm & \advccm & \uk &\cmark &  Individual (*) &  Async (*) & Strong (*)  \\ %Figures~2 \& 3 in~\cite{sana}
% \hline
\rowcolor{Gray}
\lisas~\cite{lisa}  &  \advswcm & &   & & \advccm & \xmark &\cmark &  Individual &  Async (1)& Strong  \\ %Algorithms~3\&4 in~\cite{lisa}
%% \hline
\lisaalpha~\cite{lisa} &  \advswcm & &  &  & \advccm& \xmark &\cmark& Individual &  Async& Strong   \\ %(Algorithms~1\&2 in~\cite{lisa}
%% \hline 
\rowcolor{Gray}
\seed~\cite{seed}+\seda~\cite{seda}  &  \advswcm & \advmswcm &  &  &  \advccm& \xmark & NI & Group  &  Async & Weak  \\
%% \hline
\scapi~\cite{scapi}  &  \advswcm & &   &  \advpicm &  \advccm& \skupd &\cmark& Individual/Group  &  Async& Weak  \\ %(Figures~1 \& 3 in \cite{scapi}) 
%% \hline
\rowcolor{Gray} 
  \begin{tabular}[c]{@{}l@{}}
    \erasmus~\cite{erasmus}+\\
    \lisaalpha~\cite{lisa}
  \end{tabular}
&  \advswcm & \advmswcm &   &  &  \advccm& \xmark &NI & Individual &  Async& Strong \\ %(Figure~2 in~\cite{erasmus,erasmus(extended)}) + \lisaalpha~\cite{lisa}
%% \hline 
  \begin{tabular}[c]{@{}l@{}}
    \erasmusod~\cite{erasmus(extended)} \\
    + \lisaalpha~\cite{lisa}
  \end{tabular}
& \advswcm & \advmswcm &  &  &   \advccm&\xmark&\cmark&Individual &  Async& Strong  \\ %(Figure~2 in~\cite{erasmus,erasmus(extended)}) + \lisaalpha~\cite{lisa}
%% \hline 
\rowcolor{Gray} 
\consensus~\cite{cons} & \advswcm & & & &  &\xmark&NI&Individual&Sync& Strong   \\ %Figures~1 in\cite{cons}
%% \hline 
\pads~\cite{pads}   &  \advswcm & \advmswcm &    & &  \advccm&\xmark&NI&Individual&Sync& Strong  \\ %Figure~1 in\cite{pads}
%% \hline 
\rowcolor{Gray}
\salad~\cite{salad}  &  \advswcm & &  & \advpicm & \advccm& \uk & \cmark &Individual (*)& Async (*)&Weak (*)  \\ %(Figure~1 in \cite{salad}) 
%% \hline
\wise~\cite{wise2018}  &  \advswcm & &  & &  \advccm&\xmark&\cmark&Individual &  Async& Strong    \\ %(Figure~2 in~\cite{wise2018})  
%% \hline
\rowcolor{Gray}
\slimiot~\cite{slimiot2018}  &  \advswcm & &  & \advpicm & \advccm&\nupd{}+ \att ($\dagger$) &\cmark&Individual &  Async& Weak    \\ %(Figure~1 in \cite{slimiot2018}) 
%% \hline 
\usaid~\cite{us-aid} &  \advswcm & &  &  \advpicm & \advccm&\hb&\cmark&Individual& Async& Strong    \\%(Figures~2,3, \& 4 in \cite{us-aid}) 
% \hline 
\rowcolor{Gray}
\attauth~\cite{attauth} &  \advswcm & & \advccm &  & \advccm&\puf& (\xmark)& Group& Async& Strong \\ %(Algorithm~1~\cite{attauth}) 
%% \hline
\mvs~\cite{mvs}   & \advswcm & & & &  &\xmark&---&Individual (2)& Async (2) & Strong (2) \\
\rowcolor{Gray}
\museda~\cite{museda}  & \advswcm & &  & & \advccm& \xmark&\cmark&Individual& Async& Strong    \\ %(Figure~4 in~\cite{museda}) 
%% \hline 
\sap \cite{sap}  &  \advswcm & & & & &\xmark& (\xmark)&Group &Sync&Weak \\
%% \hline 
\rowcolor{Gray}
\mtra~\cite{mtra}   &  \advswcm & &   & &  \advccm&\xmark&\cmark&Individual& Async& Weak   \\ % (Algorithms~1,2,\&3 in~\cite{mtra})
%% \hline 
\radis~\cite{radis}  &  \advswcm & &  & &  \advccm&\xmark&\cmark& %W %(Figure~8 in~\cite{radis}) 
Group (3)&Async (3)& Weak (3)  \\
%% \hline 
\rowcolor{Gray}
\esdra~\cite{esdra}  &  \advswcm & &  & &  \advccm&\xmark& (\xmark)&Individual& Async& Weak (4) \\ %(Figure~4 in~\cite{esdra}) 
% \hline 
\eapa~\cite{eapa}   &  \advswcm & &  & \advpicm & \advccm & \hb &NI&Individual& Async& Strong (5)  \\
%% \hline
\rowcolor{Gray}
\shela~\cite{shela}  &  \advswcm & &  & & \advccm&\xmark&---&Individual& Async& Weak  \\ % (Figures~1\&2 in~\cite{shela})  
 %\hline 
\healed~\cite{healed}  &  \advswcm & &  & &  \advccm&\xmark& (\xmark)&Individual& Async & Strong    \\ %(Figures~2\&4 in~\cite{healed}) 
% \hline 
\rowcolor{Gray}
\sadan~\cite{sadan}  &  \advswcm & &  && \advccm& \xmark&---&Individual& Async& Strong (6)  \\
% \hline
\sfs~\cite{sfs} &  \advswcm & &  &  \advpicm &  \advccm& \uk($\ddagger$) &\cmark&Individual (*)& Async (*)& Strong (*)   \\ %(Figures~2\&3 in~\cite{sfs}) 
% \hline
\rowcolor{Gray} 
\pasta~\cite{pasta}  &   \advswcm & &  & \advpicm & \advccm&\att &\cmark %H %(Algorithms~1,2,3 \& 4 in~\cite{pasta})  
&Individual&Sync& Weak  \\
% \hline
\dads~\cite{dads}  &  \advswcm & &  & &  \advccm&\xmark&\cmark&Individual& Async& Strong    \\ %(Algorithms~1\&2 in~\cite{dads}) 
% \hline
\rowcolor{Gray} 
\simpleplus~\cite{simple}  & \advswcm & &  & &  \advccm&\xmark& \cmark&Individual& Async&Weak   \\ %(Figures~2,4 \& 5 in~\cite{simple})
% \hline 
\cora~\cite{cora} & \advswcm & &   &&  \advccm&\xmark&\cmark&Individual/Group & Async& Weak   \\
% \hline
\rowcolor{Gray}
\sara~\cite{sara}& \advswcm & \advmswcm &   & & \advccm&\xmark&\cmark&Individual (7)& Async (7)& Strong (7) \\ % (Figure~4 in~\cite{sara})
%\hline
\fadia~\cite{fadia} & \advswcm & & & \advpicm & \advccm&\att&NI&Individual/Group&--- % (8)
                                           & Strong  \\ % (Figure~1,2,3 in~\cite{fadia})
% \hline 
\rowcolor{Gray}
\bfb~\cite{bfb} &\advswcm & \advmswcm &  & &  \advccm&\xmark&NI&Group & Async& Weak  \\
% \hline 
\mton~\cite{m2n} & \advswcm & &  && \advccm&\xmark&\cmark&Individual& Async&Strong \\
% \hline 
\rowcolor{Gray}
\swarnaagg~\cite{swarna}  & \advswcm & &  && \advccm&\xmark& \cmark&Individual& ---% (9)
                                 & Strong     \\ % (Algorithms 4,5,6  and Figure~3 in~\cite{swarna})
% \hline 
\shots~\cite{shots}  & \advswcm & &  \advccm &  &  \advccm&\puf& \cmark&Individual (*)&Sync (*)&Strong (*)  \\
% \hline 
\rowcolor{Gray}
\hola~\cite{hola} &\advswcm & &  & \advpicm & \advccm&\att{} + \hb ($\dagger$)&\cmark&Individual& Async& Strong\\
% \hline
\fesa~\cite{fesa} & \advswcm & \advmswcm&  &&   \advccm&\xmark&NI&Individual& Async& Weak  \\ % (Figure 3 in \cite{fesa}) 
% \hline
\rowcolor{Gray}
\scraps~\cite{scraps}&\advswcm & &  && \advccm&\xmark&\cmark&Individual& Async& Strong  \\ %(Figure~4 in~\cite{scraps})
\hline 
\end{tabular}
}
\end{center}
% \vspace{-15pt}
\end{table*}

% \textbf{Notes:} \quad  

% SR Suggests we could talk about this in future work 
% \jlp{I do not think this paragraph belongs at this location: }
% We consider a pre-assumption for the properties based on our earlier discussion in~\cref{subsec:mitigation-detection} for detecting physical attacks. The pre-assumption indicates that if a device is physically captured for a certain amount of time, the adversary cannot use the leaked information from that device to forge attestation results for other devices in the swarm. However, the properties are not guaranteed, but likely achievable, for CRAs that mitigates such attacks.

We now present different techniques used by CRA protocols to select
the provers to be attested (\cref{subsec:attestation set}). Then, we
present our eight security properties (\cref{def:individual
  (a)synchronous weak/strong attestation} and \cref{def:group
  (a)synchronous weak/strong attestation}) that address attestation
correctness between the provers and the relying party (see
\cref{fig:roles}).
% Our definitions are hierachical such that we first consider the
% correct status of an individual prover (or group of provers) at
% particular point in time (\cref{sec:status-correctness}). Then we
% consider when the status of the selected set of provers can be
% considered to be correct and this yields two definitions that reflect
% when the results are collected
% ($6.3) . These definitions are then used in the overall attestation
% properties which enable us to provide formal security properties that
% can be clearly cross referenced to the notion Quality of Swarm (QoSA)
% that exists within the literature.
Our definitions are hierarchical and first consider the correct
status of an individual prover at particular point in time
(\cref{sec:status-correctness}). Then we consider synchronisation
aspects, which yields two further definitions that reflect \emph{when}
a collection of provers has a particular status
(\cref{sec:synchronicity}).
% and cover attestation guarantees across {\em status correctness}
% (\cref{sec:status-correctness}) and {\em (a)synchronicity}
% (\cref{sec:synchronicity}), % , and {\em QoSA} (\cref{subsec:qosa}),
These are then combined with the notion of \emph{Quality of Swarm
  Attestation} (QoSA)~\cite{lisa} to give rise to two distinct sets % of
% four security 
properties depending on whether the CRA procedure can
determine the status of an individual prover or a group of provers as
a whole~(\cref{sec:attest-prop-1}).

Although the QoSA dimension has been extensively discussed in prior
works, our contribution is to define status correctness and
synchronicity, which lack formal definitions in the literature.

\begin{note}[\cref{table:security
    properties}]
The following provides some
  additional notes on \cref{table:security
    properties}.
  \begin{itemize}
  \item [\ccmark:] adversary model applies 
  \item [\cmark:] aims to hold
  \item [\xmark:] does not aim to hold
  \item [---:] the property is not inferable from the paper
  \item [NI:] non-interactive protocol thus \va is  not applicable 
  \item [(*):] % Since the CRA mitigates physical attacks, but does not detect them, t
    These properties are not guaranteed but it is likely to be achieved. 
  \item [($\dagger$):] As described in \cref{subsubsec:physical-detection}.
  \item [($\ddagger$):] \sfs~\cite{sfs} additionally uses a puncturable
    forward-secure signature scheme to prevent an attacker from forging
    signatures from previous attestation rounds. 
  \item [(\xmark):] These properties are conditional. Although
    \attauth claims ``mutual authentication between a prover and the
    verifier'' as a security objective, it does not prevent replay
    attacks because it does not use any known countermeasure (e.g.,
    nonce or timestamp).  \sap acknowledges that it does not handle
    replay attacks (but gives suggestions of how to change the
    protocol). \esdra and \healed use nonces but only to prevent
    replay attacks on the response, not the challenge.
  \item [(1):] \lisa's definition of synchronicity, which is different from ours, means each device waits for all
    of its children's attestation responses before submitting its own. 
  \item [(2):] For \mvs (consensus), we may adapt our definition for attestation property so that a group of untrusted verifiers collaborate to attest the swarm.
  \item [(3):] For \radis, we adapt the attestation definitions so that the verifier attests a set of services as a group. We infer from the paper that services are asynchronous and if a prover does not respond, the corresponding service is marked as unhealthy, since it does not match to unexpected healthy one. 
  \item [(4):] \esdra uses some probabilistic parameters to increase the fault tolerance against false positive reports, yet it does not claim that it prevents them all.
  \item [(5):] In \eapa, the verifier accepts the absence message of
    healthy devices and deletes the suspicious
    devices in absence message from the swarm topology.
  \item [(6):] For \sadan, we may adapt our definition for attestation property so that the verifier will be a random jury of untrusted devices.
  \item [(7):] \sara's attestation target is service flow plus program memory. We may adapt our attestation definitions like the way we do for \radis.
  \end{itemize}
\end{note}

% Status correctness captures false positive and false negative
% attestation results.

% \sa{% Unlike previous works with the limitation of having ad hoc definitions for security properties, this work introduces a novel unified approach for comparing CRA protocols.

% %We are not specifically show how the designer can use these in CRA development.
% %Specific use-cases for the protocols  e.g., where would you need this? but Papers usually do not  go to the detail of the applications.
% %We are not sure when do we need the CRA to by synchronous
% %Individual/Group is already done.
% %Talking about the complexity / resources / memory / …. E.g, weak needs shorter messages
% }

% I don't mind the compact intro but maybe we could expand to something like because the hieratchical nature is not coming  through here. "  

% {\em Individual/Group
%   (A)Synchronous Weak/Strong Attestation}. 

\subsection{Attestation Set}
\label{subsec:attestation set}

When initiating the attestation procedure, there are several options
for selecting the {\em attestation set}, i.e., the set of provers to
be attested. Ideally, one would only attest provers when strictly
necessary since attestation itself prevents the prover devices from
performing useful work. Attestation may have significant performance
implications in resource-constrained devices and settings where the
prover devices execute time-sensitive operations.

Most of the CRA protocols studied in this paper attest \emph{all} of
the provers. However, to improve scalability, the set of all provers
may be partitioned into different classes~\cite{slimiot2018}, e.g.,
based on geographical locations in static networks, common tasks among
devices, network traffic distribution, etc.

Some protocols only attest a {\em sample} of provers, using various
techniques to select the attestation set. One option is to simply pick
a random subset of provers, e.g., as in \attauth~\cite{attauth}) and
\healed~\cite{healed}.  More sophisticated CRA designs such as
\wise~\cite{wise2018} use AI-based classification techniques over
features such as the maximum amount of time between attestation
requests and prover features such as its geographical location and
underlying hardware. In \esdra~\cite{esdra}, cluster heads assign and
update reputation scores on the provers. The protocol attests provers
when their reputation drops below a certain threshold. In
\sadan~\cite{sadan} each prover monitors other provers in the network
and triggers attestation on any prover that it finds suspicious.
\wise~\cite{wise2018} provides probabilistic guarantees about the
healthiness of all provers in the network even after sampling a subset
of provers, where the history of previous attestation periods and
device characteristics are used to select the attestation set.

In the remainder of this section, we define attestation properties
between the prover(s) selected for attestation (irrespective of the
selection strategy) and the relying party.

% \item [{\em Selected}] protocols, where a verifier selects a subset of
% provers to perform attestation. For example, after the CRA protocol
% \sara~\cite{sara} has 
% %asynchronously 
% collected historical data of
% the services in an %large-scale 
% IoT system, the verifier only selects
% the actuators, which perform the final action, to perform
% attestation. In this way, the verifier can build a trust in all the
% devices involved in the provision of that specific service.
% \end{description}

\subsection{Status Correctness}
\label{sec:status-correctness}
A CRA protocol must maintain a list of states that are 
considered acceptable for each prover. This list may be updated over time, e.g., to reflect new software versions. 

If a prover's state at some time is in the list of acceptable states for that prover at that time, then the prover has a \emph{valid state} at that time and is considered \emph{healthy}. If not, the prover is \emph{unhealthy}. The condition of being healthy or unhealthy is the prover's \emph{status}.

In some protocols, each status (i.e., $\healthy$ or $\unhealthy$)
observed by the relying party must match the actual state of the
associated prover. Other protocols only require this for the
$\healthy$ case, allowing false-positive results. This gives rise to a
weak and a strong version of status correctness.

\begin{definition}[Weak Status Correctness]\label{def:weak status correctness} 
We say that $s$ is a weakly correct status for prover $p$ at time $t$ iff, whenever $s=\healthy$, $p$ had a valid state at $t$. 
\end{definition}

\begin{definition}[Strong Status Correctness]\label{def:strong status correctness}
We say that $s$ is a strongly correct status for a prover $p$ at time $t$ iff, 
(1) it is a weakly correct status for $p$ at $t$, and
(2) whenever $s=\unhealthy$, $p$ did \emph{not} have a valid state at $t$. 
\end{definition}

\subsection{Synchronicity}
\label{sec:synchronicity}

(A)synchronicity enables the categorisation of CRA protocols based on
their ability to compute provers' measurements at different or the
same time in an interval. The execution of CRA protocols proceeds over
a period of time often between devices with no real-time
clocks~\cite{simple,seda,sana,sara}, or only loosely synchronised
ones~\cite{slimiot2018,us-aid,pasta}. This means that we need to be
able to determine the correctness of a set of provers depending on
when the results from the provers are collected. A verifier typically
knows the result was produced within a known time interval $T$ but it
does not know at exactly what time the result was generated.

%\jlp{I dislike this paragraph and want to remove it. Talking about $G$ is confusing as $G$ is not used anywhere. It is possible that a higher level discussion about the set being attested would be helpful, but not this:}
%In CRAs, a verifier attest a number of provers. We denote the set of all provers in the swarm by $G$ 
%and the the set of provers that the verifier makes a decision about by $P$, which may be equal to or a subset of $G$ (e.g., a decision may not be reached for unresponsive provers). Note that a verifier may make a decision about any prover in $G$ (including unresponsive provers), e.g., a verifier may classify unresponsive provers as unhealthy in $P$.

\begin{definition}[Asynchronous Weak/Strong Correctness]\label{def:asynchronous weak/strong correctness} 
The statuses for a set of provers $P$ are asynchronously weakly/strongly correct in $T$ iff, for all $p \in P$ there is some time $t \in T$ where the decided status of $p$ is weakly/strongly correct for $p$ at $t$. 
\end{definition}

According to this definition, the state of each prover in $P$ can be measured at a different time in $T$ -- there is no guarantee that all provers are ever simultaneously healthy, even if the decided statuses of all provers in $P$ are \healthy. If malware were to hop between devices during $T$ it would be possible for every prover in the swarm to be correctly viewed as $\healthy$ in $T$, even though at no single time in $T$ is every prover in $P$ $\healthy$~\cite{rata}. 

The stronger \emph{synchronous} correctness property avoids this limitation,  requiring all statuses to be correct at a single time $t \in T$. 

\begin{definition}[Synchronous (Weak/Strong) Correctness]\label{def:synchronous weak/strong correctness}
The statuses for a set of provers $P$ are synchronously weakly/strongly correct in $T$ iff, there is some time $t \in T$ such that for all $p \in P$ the status of $p$ is weakly/strongly correct for $p$ at $t$. 
\end{definition}

% \subsection{QoSA (Quality of Swarm Attestation)}\label{subsec:qosa}

% The
% drawback of L-QoSA is when provers have limited memory. In such cases,
% a hierarchy protocol which utilises edge verifiers
% Finally, in a protocol with Full-QoSA, e.g., a version of
% \darpa~\cite{darpa}
% %(Figures 1, 2, and 3~\cite{darpa}) 
% with .
% \sa{add explanation about reporting healthy, unhealthy, both, if from unhealthy ones healthy ones can be inferred, and if from healthy ones unhealthy ones can be inferred. All of these are considered to be L.}

%DARPA is B is mentioned directly in shela paper
%We assume that the
%edge verifiers have sufficient storage to keep track of the attestation status of each swarm node, ensuring the highest level
%of QoSA

\subsection{Attestation properties}
\label{sec:attest-prop-1}

Information about the provers' software states may be collected in
several ways. Carpent \etal~\cite{lisa} define the notion of a
\emph{Quality of Swarm Attestation (QoSA)}, which describes how much
information a relying party is able to retrieve from a report it
receives (see~\cref{table:categories}). QoSA can be used to
distinguish protocols as follows\footnote{Carpent et al.~\cite{lisa}
  also define the notion of a {\em Full QoSA}, where a report
  identifies a list of attested devices along with their connectivity,
  i.e., swarm topology. However, as far as we are aware, there is no
  CRA protocol that implements Full QoSA. }.
\begin{description}[itemsep=2pt,leftmargin=0pt]
\item [{\em Binary QoSA} (B-QoSA)] protocols, where the report only
  allows the relying party to distinguish between two states: all
  provers are healthy, or some prover is
  unhealthy. % in which case the whole
  % swarm is labelled unhealthy.
\item [{\em List QoSA} (L-QoSA)] protocols, where report only allows
  the relying party to identify the healthy or unhealthy provers, or
  both.
\item [{\em Intermediate QoSA} (I-QoSA)] protocols, which lie between
  B-QoSA and L-QoSA.  Here the provers are partitioned into sets of
  provers, and the report allows the relying party to identify the
  healthy sets without distinguishing the individual provers within
  the set.
%\item [Full-QoSA] protocols, where a report identifies a list of attested devices along with their connectivity, i.e., swarm topology.

% a list of healthy sets
% without % the report that verifier knows, and
% % the verifier registers a binary report for each subset \bd{I don't
% % quite understand this... }
%we cam omit Full-QoSA, since it as extension of QoSA only used in one protocol}

% \item [Full-QoSA] \bd{Full-QoSA, the swarm topology is also attested}
% \bd{I don't understand what Full QoSA is. What is the distinction
% between ``what is attested''? }
\end{description}
%Clearly, increasing the collected information about the swarm 
%increases both communication and computation overhead. However, in
%most applications it is necessary to locate unhealthy provers so that
%further steps such restoring the healthy code on such provers (e.g.,
%\healed~\cite{healed}) or disconnecting them from their neighbouring provers can
%be performed (e.g., \usaid~\cite{us-aid}). 
\noindent As summarised in \cref{table:categories}, the majority of
CRA protocols are L-QoSA, since in most applications, it is necessary
to precisely locate unhealthy provers to take further action, e.g.,
restoring healthy code~\cite{healed} or disconnecting them from their
neighbouring provers~\cite{us-aid}.  Furthermore, some CRA protocols
may have more than one QoSA, e.g., \cora~\cite{cora} is initially a
Binary protocol and if the result of the attestation is 1, it
concludes that the entire swarm is healthy, otherwise, the protocol
concludes that at least one device is unhealthy and proceeds to the
detection phase where \cora is List-QoSA. In \scapi~\cite{scapi}, the
QoSA metric is a parameter of the protocol, referred to as {\em
  attestation type}, whose value can be either Binary or List, and the
relying party decides on the value of this parameter. \fadia takes a parameter for how many measurements should be aggregated. This number can be 0 (L-QoSA), greater than the number of provers being attested (B-QoSA), or an intermediate value (I-QoSA).

QoSA gives rise to two different ways for performing CRA: the protocol
can deliver to the relying party the status of each prover in the
swarm \emph{individually}; or the relying party can be given one or
more statuses, each representing the trustworthiness of a \emph{group}
of provers. We define security properties for each type of
assessment. In these definitions we use the term \emph{run}, which is
a single execution of a protocol role by a device.

\subsubsection{Individual Attestation}\label{subsubsec:individual attestation}
Individual Attestation is the simplest form of CRA, where the relying
party receives one status per prover. This corresponds to L-QoSA.

\begin{definition}[Individual (A)Sync. Weak/Strong
  Att.] \label{def:individual (a)synchronous weak/strong attestation}
%  A protocol guarantees Individual (A)Synchronous Weak/Strong Attestation iff, whenever the relying party completes a run of the protocol apparently with a set of provers $P$, believing that each $p_i \in P$ has some status $s_i$ during some time interval $T$, then the statuses for $P$ are (a)synchronously weakly/strongly correct at $T$.
  A protocol ensures Individual (A)Synchronous Weak/Strong
  Attestation to a relying party for a set of provers $P$ during a time interval $T$ iff, whenever the relying party completes a run of the
  protocol believing that each $p_i \in P$ has some status $s_i$
  during the time interval $T$, then the statuses for $P$ are (a)synchronously weakly/strongly correct in $T$.
\end{definition}

While in many protocols $P$ is decided when attestation is initiated (corresponding to the attestation set in \cref{subsec:attestation set}), our property is flexible and only requires it to be fixed at the end of the protocol run, when the claim is made. In PADS~\cite{pads} for example, allows some provers in the attestation set to have an ''unknown'' status. In this case, $P$ will contain only the provers that are reported to be healthy or unhealthy. 

\subsubsection{Group Attestation}\label{subsubsec:group attestation}
In B-QoSA and I-QoSA protocols, a set of provers can have a \emph{group status} -- a single healthy/unhealthy value assigned to a set of provers as a group, based on the state of each prover in that set. 

\begin{definition}[Group (A)Sync. Weak/Strong
  Att.]\label{def:group (a)synchronous weak/strong attestation}
  % Suppose a relying party completes a run of a protocol apparently
  % with a set of provers $P$, believing that during some time interval
  % $T$ one or more sets of provers $G_i \subseteq P$ each have some
  % group status $s_i$. The protocol guarantees
  Let $G_1,\ldots,G_n \subseteq P$ for $n\geq 1$ be one or more sets of provers. Suppose a relying party completes a run of a protocol obtaining, for a time interval $T$, the statuses $s_1, \dots, s_n$ for the sets of provers $G_1,\ldots, G_n$, respectively. Then the protocol guarantees
  \begin{itemize}[itemsep=0pt,leftmargin=15pt]
      \item {\em weak asynchronous group attestation} iff, for all group statuses $s_i$, if $s_i=\healthy$ then for all $p \in G_i$ there is a time $t$ in $T$ where $p$ had a valid state at $t$,
      \item {\em weak synchronous group attestation} iff there is a time $t \in T$ such that, for all group statuses $s_i$, if $s_i=\healthy$ then for all $p \in G_i$, $p$ had a valid state at $t$,  
      \item {\em strong (a)synchronous group attestation} iff (1) it guarantees weak (a)synchronous group attestation and (2), for all group statuses $s_i$, if $s_i=\unhealthy$ then at some time $t$ in $T$, at least one $p \in G_i$ had an invalid state.
  \end{itemize}
\end{definition}

% \jlp{I do not think this paragraph makes sense in the context of a relying party. It could be rewritten but I am wondering if it might be easier (and save space) if we delete it entirely. } \jlp{Yes. Make it clear that $P$ is decided at the end of the protocol}
% When attesting provers as a group, the verifier typically makes a decision about the health of the group it initially aimed to attest, regardless of whether or not all of those provers participated in the protocol. However, our definition is flexible and allows a verifier to make decisions about a subset of the provers it attests -- $P$ is not fixed until the end of the run.   

Existing CRA protocols predominantly define a group to be unhealthy if
one or more provers in that group have an invalid state and our
property follows this. The one exception among the CRAs
%CRA protocols
included in this paper is \bfb~\cite{bfb}, which allows more than one
prover (up to some threshold) to have an invalid state, but still for the
group to be considered healthy. \cref{def:group (a)synchronous
  weak/strong attestation} can be relaxed to allow one or more provers
not to be in a valid state for all times in $T$. 
% to not have a time $t$ in $T$ where it has a valid state. 
We assume
this modified version of \cref{def:group (a)synchronous
  weak/strong attestation} in \cref{table:security properties}.
We also note that for B-QoSA protocols it is possible to simplify this property as there is only one group $G_1 = P$.

%% file: validation.tex
\section{Examples and
  Validation} %\label{sec:security properties}
\label{sec:examples-validation}

In \cref{subsec:security properties examples}, we present particular
CRA protocols that exemplify the security properties that we proposed
in \cref{subsec:verifier authentication} and \cref{sec:security
  properties} and map to these protocols in \cref{table:security properties}.
In \cref{subsec:simpleplus} we focus on one of the examples, \simpleplus~\cite{simple}, and validate the
implementability of our security properties in a formal analysis of
\simpleplus with respect to the properties.
%Then in \cref{subsec:validation}, we validate
%implementability of our security properties by using the \simpleplus
%protocol as an example against its stated properties in
%\cref{table:security properties}.

\subsection{Security Properties Examples}\label{subsec:security properties examples}

% \cref{def:verifier authentication} defines an authentication property
% between the initiator and the provers.  

\newcommand{\msat}{aim to satisfy{} }
\newcommand{\asat}{aims to satisfy{} }

In the following descriptions (and in \cref{table:security
  properties}), we say that the considered CRA protocols \emph{\msat}
(rather than satisfy) our security properties for two reasons. First,
while our properties are derived from the reviewed protocols' security
goals, except for \simpleplus, we have no proof that a considered
protocol's claimed security property implies our security
property. Second, most of the reviewed CRA protocols use proof
sketches or informal reasoning to justify their security claims.
Note that the attestation properties
stated in \cref{def:individual (a)synchronous weak/strong
  attestation,def:group (a)synchronous weak/strong attestation}
encompass eight types of attestation correctness properties. However,
we offer a concrete CRA example for only six of them %here
because %to the best of our knowledge, there are currently
none of the %40 CRA 
protocols we study address Group Asynchronous Strong or Group Synchronous Strong
attestation. %(GSS)
%or group asynchronous atrong attestation. %(GAS). 
% Additionally, we present an example that ensures verifier authentication. 

\begin{description}[itemsep=2pt,leftmargin=0pt]
\item [{\em Initiator Authentication} (IA).]  \cora~\cite{cora} \asat
  \va under \advsw and \advc by using an algebraic message
  authentication code scheme with a counter on each device, which
  is % initialised to 0, and
  incremented by the verifier and each prover after each attestation
  procedure. The counters are stored in the trusted environment of all
  devices and are used to monitor the attestation sequence number,
  preventing replay attacks. % The counter values are
  % stored locally in the trsuted environment of all devices.

\item [{\em Individual Asynchronous Weak Attestation} (IAW).]
  \
  
  \simpleplus~\cite{simple} \asat IAW under \advsw and \advc. Here, all devices have shared attestation and
  authentication keys secured within their Trusted Computing Module
  (TCM), isolated by a \microvisor. The TCM also safeguards
  attestation code and a counter for report freshness. Provers
  calculate measurements asynchronously using the attestation key and
  generate reports, which include the counter and are signed with the
  authentication key. These reports consist of $1$-bit binary values,
  and the final aggregated report comprises $n$-bit values. A `$1$' in
  the $i$-th bit indicates a healthy status for $i$-th prover, while
  `$0$' denotes unhealthiness. Thus, the verifier can determine the
  status of the individual provers contributing to the
  report. Unresponsive provers are considered to be unhealthy.

\item [{\em Individual Asynchronous Strong Attestation} (IAS).]\
  
\usaid~\cite{us-aid} \asat IAS under \advsw, \advc and \advpi. 
It operates without a trusted centralised verifier, relies on network owner providing devices with attestation details, and secures protection through \smart~\cite{smart} or \trustlite~\cite{trustlite} architectures. \usaid detects absence using DARPA's heartbeat protocol, with neighbouring devices mutually authenticating through encrypted messages and checking each other's measurements asynchronously. Successful attestations result in reports stored in separate lists for healthy and unhealthy devices.

\item [{\em Individual Synchronous Weak Attestation} (ISW).]
  \pasta~\cite{pasta}, already explained in
  \cref{subsec:mitigation-detection}, \asat ISW under \advsw, \advc
  and \advpi.
\item [{\em Individual Synchronous Strong Attestation} (ISS).]
  \pads~\cite{pads}
  aims to satisfy ISS under \advsw, \advmsw, and \advc.  Here,
  each prover % simultaneously
  performs a local attestation through its
  internal %Trusted Execution Environment (TEE)
  TEE. %  to validate its measurement against an expected healthy
  % one
  Periodically, each prover shares its observations representing the
  statuses of provers in the swarm. Other provers receiving these
  messages verify their authenticity, check their validity within a
  specified time interval, and use a consensus algorithm to share
  their knowledge with the swarm. Each prover keeps track of
  responsive unhealthy, responsive healthy, and unresponsive
  provers. Over time, the view of every prover converges to the true
  state of the swarm, allowing a verifier to query a final report from
  any of them. % The verifier checks the report's authenticity and
%   ensures it falls within a specified time interval.
% %the maximum time interval between consecutive attestation runs. 
% If any checks fail, it outputs $0$; otherwise, it outputs $1$ and records healthy and unhealthy devices.

\item [{\em Group Asynchronous Weak Attestation} (GAW).]
  \seda~\cite{seda}  aims to satisfy GAW under \advsw and \advc.
It employs a spanning tree topology where parent devices have separate shared keys with their child devices. Attestation results are generated asynchronously. When a parent attests its children, each child sends a report signed with the shared key stored in its ROM, protected by %a Memory Protection Unit (MPU)
an MPU. % Freshness is maintained using a session key also protected by the MPU. 
The final aggregated report % in \seda
includes information about the number of successfully attested provers %($\beta$) 
and the total number of provers attested %($\tau$) 
in the subtree rooted at the initial prover.
A verifier, aware of the total number of provers ($s$), checks if %$\beta$ and $\tau$ 
these two are equal to $s-1$ and if the status of the initial prover is healthy. If these conditions are met, the verifier determines the overall group health status as healthy, unhealthy otherwise. Unresponsive provers are assumed to be unhealthy in this protocol.

\item [{\em Group Synchronous Weak Attestation} (GSW).]
  \sap~\cite{sap} \asat GSW under \advsw.  It employs a balanced
  binary tree topology where the root is the verifier. Attestation
  requests propagate through the tree, and secure clocks on devices
  ensure proper verification timing. Each device, upon receiving the
  request, performs attestation and sends the result to its
  parent. The parent combines the received results with its own
  attestation result %using XOR
  and forwards an aggregated report to its parent.
%When the network-wide attestation completes, 
Eventually, the verifier receives the final %aggregated 
report and validates it. The binary value in the report indicates whether the entire swarm is trustworthy (all provers are healthy) or not (one or more provers are unhealthy or unresponsive).
%\sara~\cite{sara} guarantees initiator authentication under software, mobile software, and Dolev-Yao adversaries. Here, the verifier initiates authentication with a publisher (prover) by sending a signed challenge using its private key. The publisher verifies the request using the verifier's public key and proceeds with the attestation process only if the verification is successful.

%Three example CRA protocols that guarantee \va are \mtra~\cite{mtra}, \simpleplus~\cite{simple},

%\mtra~\cite{mtra} guarantees \va using a one-way hash chain where the last hash key ($V_k$)
%(i.e., the committed value of hash chain, $V_k$)
%is distributed to all devices. $V_{k_i}$ is to authenticate the $(i + 1)$th
%challenge from the verifier. Only the base-station (trusted verifier) keeps all the
%elements in this hash chain. 
%The number of key chain elements should be sufficient to
%cover the lifetime of the IoT network.  
%The committed element of the
%one-way key chain can be made known to public as only the base-station has previous key for key verification.

\end{description}

\subsection{Encoding of Properties}\label{subsec:encoding}
We encode our security properties: IAS, IAW, GAS and GAW, as generic \tamarin lemmas that are independent of any particular protocol. The \tamarin lemma encoding IAS and IAW is shown in~\cref{fig:individual-async-attestation}, while the Group Asynchronous Attestation lemmas can be found
in~\cref{fig:async-group-att-strong}.

\begin{figure}
  \begin{lstlisting}
All V P0 P1 att0 att1 cntr #att_end . 
Attestation_Complete_I(V, P0, P1, att0, att1, cntr) @ 
#att_end ==> Ex #att_start . (
 Attestation_Start(V, P0, P1, cntr) @ #att_start 
& (att0=healthy ==> IsHealthy(P0, #att_start))
& (att0=unhealthy ==> IsNotHealthy(P0, #att_end))//Strong
& (att1=healthy ==> IsHealthy(P1, #att_start))
& (att1=unhealthy ==> IsNotHealthy(P1, #att_end)))//Strong
\end{lstlisting}
% \vspace{-5pt}
\caption{Lemma for IAS. The Lemma for IAW is obtained by omitting
  the % the $attx = unhealthy$
  cases (marked {\color{green!60!black}//Strong})}
  \label{fig:individual-async-attestation}
  % \vspace{-15pt}
\end{figure}

In the \advsw adversary
model, once a prover becomes unhealthy, it can never become healthy
again. Therefore, if a prover is healthy at any time in an interval,
it must be healthy at the start of the interval. Conversely, if a
prover is unhealthy at any time in an interval, it must be unhealthy
at the end of the interval. We exploit this to limit the possible
values of $t$ to \attstart or \attend, depending on whether the
observed status is $healthy$ or $unhealthy$, respectively. \attstart
is tied to when the verifier first sends an attest request message for
a round (each round is uniquely identified by a monotonically
increasing counter {\tt cntr}), while \attend denotes the verifier
{\tt V} completing the attestation protocol and deciding on the status
of the provers ({\tt P0} and {\tt P1}). {\tt IsNotHealthy} and {\tt
  IsHealthy} are predicates that check whether the state of the prover has been modified.

\subsection{Verification of \simpleplus}\label{subsec:simpleplus}
We analyse the \simpleplus protocol using the lemmas encoded
in~\cref{subsec:encoding},
% We use the our security properties by 
showing that \simpleplus
satisfies Individual Weak Asynchronous Attestation and Group Weak
Asynchronous Attestation, but fails to satisfy the Strong variants of
these properties under the \advsw and \advc adversary
models.\footnote{Our models are submitted as supplementary material 
 and will be made available.} 
% ~\cite{TamarinSIMPLE}.}
We build on Le-Papin
\etal's~\cite{Jay} symbolic protocol analysis of \simpleplus using
\tamarin~\cite{tamarin}, which was the first symbolic analysis of a
CRA protocol. In particular, we reuse Le-Papin \etal's \tamarin model
of \simpleplus, implement our security properties within this model,
then verify them automatically.  Like Le-Papin \etal, our analysis
considers a scenario comprising two provers and up to two rounds. This
extension shows that our properties can be used in the formal analysis
of CRA protocols.
A more detailed
description of Le-Papin et al.'s model, can be found
in~\cref{app:tamarin}.

% We expect that the other attestation properties can be
% similarly encoded as \tamarin lemmas but leave this for future work.

% \jlp{The following four paragraphs are newly added. I am not convinced it's necessary but this is, as I understand it, what I was asked to write:}

% A physical attack on a prover is modelled by its keys being revealed
% to \advc, \bd{Next part a bit unclear}  The DY adversary then models both keys being read and used
% elsewhere or being used by an attacker on the prover.

% The model is bounded with two provers and up to two rounds. Concurrent execution of the CRA protocol by multiple swarms is not considered.

%%% Local Variables:
%%% mode: latex
%%% TeX-master: "main"
%%% End:

%% file: conclusion.tex
% \section{Related Work}
% \label{sec:related-work}

\section{Discussion and 
  Future Work}\label{sec:conclusion}
The \fw framework proposed in this paper offers protocol designers
guidance on tailoring CRA protocols to specific needs. For instance,
if the goal is to minimise false positive attestation results,
designers should opt for a CRA ensuring \emph{strong status
  correctness}. Similarly, if the aim is to reduce complexity by
exchanging shorter attestation messages, a CRA evaluating the status
of a \emph{group} of provers collectively is preferable. Additionally,
in scenarios where adversaries may evade detection by moving between
provers, a promising choice is a CRA guaranteeing a \emph{synchronous}
security property, showcasing the practical applications and benefits
of \fw's unified approach.

Examining \cref{table:security properties}, we find that the majority
of CRA protocols ensure Individual Asynchronous Weak
Attestation. Three primary reasons contribute to this trend.
\begin{enumerate*}[label=\bfseries(\arabic*)]
\item CRA designers prefer to enable relying parties to take
  additional steps, such as healing or disconnecting unhealthy provers
  from the rest of the swarm, necessitating individual
  detection~\cite{healed}.
\item Achieving synchronicity is
challenging, especially in dynamic environments. While it is great in
theory, it is difficult to achieve in practice as it requires
strong clock synchronisation (such as in \sap~\cite{sap}) or dedicated
hardware (such as in \rata~\cite{rata}). As shown
in~\cref{table:security properties}, almost all CRA protocols only aim
for asynchronous correctness.
\item The primary
concern for relying parties is ensuring that if an attestation report
indicates a prover's health at a specific point, the prover was
genuinely healthy at that time, and the concern is not avoiding false
positives---if an attestation report indicates a prover is unhealthy, 
% at a specific point, 
but the prover was genuinely healthy at that
time.
\end{enumerate*}

% \paragraph{Research challenges} 
By developing our \fw framework we
identify several possible research challenges. We split these into three main themes. 
% A promising avenue for future research involves exploring the
% contextual aspects of CRA design. Understanding the context in which
% these protocols are created can provide valuable perspectives for
% enhancing and refining metrics and security properties in future CRA
% designs.  Furthermore, t
\begin{description}[itemsep=3pt,leftmargin=0pt]
\item[\em Protocol robustness.] The security properties proposed in this
  work are specifically applicable to CRAs where attestation failure
  may be attributed to adversarial attacks or network delays. However,
  it is acknowledged that attestation failures can also occur due to
  bugs in an implementation. Addressing this presents a research
  challenge, exploring the possibility of verifying that attestation
  is bug-free, particularly in swarms with multiple devices where
  multiple states may need resetting. Additionally, certain CRA
  protocols assume bug-free attestation code, posing a vulnerability
  and suggesting another area for research within the community.
  
\item [\em Formal verification.] In \cref{subsec:encoding} and \cref{subsec:simpleplus}, we have
  shown that \fw's security properties are amenable to formal
  verification. The next steps for future work are to formally prove
  that the security properties shown in \cref{table:security
    properties} are indeed satisfied by the CRA protocols. We
  anticipate that this may require significant effort since the
  assumptions made by the protocols must be faithfully
  represented. Given the complexity of many of the protocols, there
  are also scalability challenges that must be overcome.
  % , since they
  % have not been verified with respect to their aimed properties. For
  % many of the 40 protocols, this will constitute their first formal
  % analysis and either provide assurance for their correctness or
  % uncover design flaws.
\item [\em Protocol design.]  Tables 1 and 2 highlight numerous
  unexplored CRA protocol possibilities, such as protocols that
  guarantee group (a)synchronous strong attestation. Other
  possibilities include use of special hardware such as DICE (Device
  Identifier Composition Engine), which is a security control for
  assuring the trustworthiness of IoT devices.
  % and the software, for
  % attestation. DICE was created by the (TCG) Trusted Computing Group
  % as an alternative for or addition to a TPM, and its purpose is to
  % derive a cryptographic identity from a device's firmware and a UDS
  % (Unique Device Secret). Based on this cryptographic identity, the
  % firmware can derive keys for attestation and other purposes. They
  % are also interesting, since their adversary model match adversary
  % capabilities to the full spectrum of physical properties (and
  % physical attackers) to logical properties (and software bugs and
  % adversaries exploiting them).  
  There are RA protocols that use TCG DICE~\cite{rolling, disclosure,
    delegated} so an interesting line of research is to find out if
  these can be used to
  implement % can also be fit in our \fw model for
  CRA protocols.  Another promising avenue for future research is to
  investigate if the efficiency trick of \cora~\cite{cora} can be
  extended to other CRA protocols. This way, in the common case where
  all devices are healthy, the protocol just reports a simple healthy
  statement and it only goes into per-device status if at least one
  device is unhealthy.  % \jlp{For
% \item [\em Applications.] While CRA protocols have received much
%   academic attention, they are yet to be thoroughly tested on real
%   applications.

\end{description}

%% file: appendix.tex
 \newpage%\onecolumn
\section{Underlying hybrid environment}\label{app:architecture}

The underlying hybrid environment for specific implementations in the CRA papers are shown in~\cref{table:hw}.
This provides a useful summary of the different environments and shows
that TrustLite is currently the most popular which is applicable across all types of protocols.

\begin{table}[!t]
\scriptsize
\begin{center}
\centering\caption{Underlying hybrid environment for specific implementations in the CRA papers} 
\label{table:hw}
\begin{tabular}{|p{0.31\linewidth}|p{0.62\linewidth}|}
  \hline 
\rowcolor{DarkGray}
     architecture & protocol name\\
  \hline 
   ARM TrustZone &    \shela~\cite{shela}, \radis~\cite{radis}  \\
 \hline 
   \hydra&   \erasmus\cite{erasmus}+\lisaalpha\cite{lisa}, \fesa\cite{fesa}  \\
\hline  
   \smart&  \darpa\cite{darpa}, \healed\cite{healed}, \sana\cite{sana}, \seda\cite{seda},  \seed\cite{seed}+\seda\cite{seda}, \sfs\cite{sfs}, \usaid\cite{us-aid}  \\ \hline
   \smartplus& \erasmus\cite{erasmus}+\lisaalpha\cite{lisa},  \lisaalpha\cite{lisa}, \lisas\cite{lisa}   \\ \hline
   Software+PUF &   \shots~\cite{shots}, \usaid~\cite{us-aid} \\
\hline 
   \trustlite&  \bfb\cite{bfb}, \dads\cite{dads}, \eapa\cite{eapa},   \esdra\cite{esdra},  \healed\cite{healed},  \pads\cite{pads}, \sana\cite{sana}, \sap\cite{sap}, \seda\cite{seda},  \seed\cite{seed}+\seda\cite{seda}, \usaid\cite{us-aid} \\ \hline
   \tytan&  \sana~\cite{sana} \\ \hline
\rowcolor{DarkGray}
     hardware & protocol name\\
  \hline 
   Arduino &  \slimiot~\cite{slimiot2018},  \wise~\cite{wise2018} \\
\hline  
   ATmega1284P-Xplained & \scraps~\cite{scraps}\\ 
\hline
   ESP32-PICO-KIT V4 & \pasta~\cite{pasta} \\
\hline  
   IntelN5000 &  \sadan~\cite{sadan} \\
\hline  
   LPC55S69-EVK & \scraps~\cite{scraps}\\ 
\hline
   MicroPnP & \slimiot~\cite{slimiot2018}, \wise~\cite{wise2018}  \\
\hline   
   Microship STK 600 %with AVR Atmega644P
& \wise~\cite{wise2018}  \\
\hline  
    Odroid XU4 &  \mtra~\cite{mtra} \\
\hline
   Raspberry Pi & \esdra~\cite{esdra},\fadia~\cite{fadia},\healed~\cite{healed},\hola~\cite{hola},
\mtra~\cite{mtra},\mvs~\cite{mvs},\usaid~\cite{us-aid},\wise~\cite{wise2018} \\
\hline  
Raspberry Pi + PUF & \shots~\cite{shots}\\
  \hline
   Stellaris EK-LM4F120XL&  \scapi~\cite{scapi} \\
\hline
   Stellaris LM4F120H5QR&  \salad~\cite{salad} \\ 
\hline
   Teensy 3.2&  \cora~\cite{cora} \\ 
\hline  
   Tmote Sky&   \fadia~\cite{fadia}, \sara~\cite{sara} \\ 
\hline
   Xilinx FPGA &  \shela~\cite{shela} \\
\hline 
\end{tabular}
\end{center}
\end{table}

\section{Additional Metrics}\label{app:more metrics}

The metrics referred to in this appendix have been identified in the
previous literature~\cite{Ambrosin,lisa}, but their systematic
inclusion here serves to contextualise existing knowledge to show that
there is little correspondence between the strength of the security
properties of a protocol and how they align to particular network
topologies, dynamic configurations of devices and considerations of
how the final attestation report is recorded.

\subsection{Network Topology}
\label{subsec:topology}
Depending on the application, there may be many different (logical)
arrangements of devices. The network topology assumed by a
CRA protocol can be constrained by several factors, e.g., the
communication and computation capability of devices, the
environment in which the protocol is used, the adversary capabilities, and
the need to support device mobility. 
CRA protocols may
assume any of the following topologies
(shown in Column 2 of~\cref{table:categories-OLD}).

\begin{description}[itemsep=0pt,
leftmargin=0pt]
\item [Spanning-tree (\TopST)] protocols maintain a structure where the root of
the tree is the (set of trusted or untrusted) verifier(s) and the
leaf nodes are provers. Devices corresponding to intermediate nodes
may take the role of a prover or aggregator, or both. We distinguish
between the following, which are specialised forms of spanning-tree
topologies: (1) {\em Aggregation Layer (\TopAgg)} protocols, where all devices
in the intermediate nodes are aggregators and (2) {\em Balanced Binary Tree (\TopBBT)}
topologies that maintain a balanced binary tree to ensure that the
distance from the verifier (root) to the provers (leaves) is
logarithmic with respect to the number of devices in the swarm.

% \bd{This is obvious}
% In this topology, the root verifier broadcasts an
% attestation request to its children and the children rebroadcast
% the messages to their children and so on until the message is received
% by the leaf nodes. Then every node aggregates the attestation
% responses of its own subtree and sends it to its parent until the
% final aggregated message is received by the verifier.

% \item [] which has a spanning-tree topology, but the
% verifier communicates with a set of {\em edge verifiers} through a
% synchronisation mechanism. Each edge verifier attests the devices in
% a subtree whose root is the edge device. Edge verifiers may also
% exchange information about the attestation with other edge devices,
% allowing the root verifier to learn about the integrity of any in
% the swarm from any one of the edge
% devices. % An example of a swarm attestation
% % protocol with hierarchical graph topology is \shela~\cite{shela}
% % where there is an additional edge layer in between the root verifier
% % and the swarm provers of a spanning-tree topology. 
% The edge verifiers may be geographically spread, and typically have
% more computational power and storage capacity than the provers. The
% subtree corresponding to each edge verifier may run a different
% swarm attestation protocol.

\item [Distributed] protocols allow devices to communicate with many
neighbouring devices via an interconnected network.

\item [Cluster] protocols partition connected devices into
clusters (aka classes) that are periodically
attested. The clustering is based on various factors
(e.g., geographical locations in static networks, common tasks among
devices, network traffic distribution, etc.). Devices in each cluster interact with a cluster head device, and cluster heads may interact among themselves and/or with a centralised verifier.
% (see \cref{subsec:centralised}).

\item [Hierarchy] protocols maintain different layers where each layer
contains several \emph{clusters} of devices with a set of cluster
heads (aka edge verifiers). The provers in each cluster can be
attested using (potentially different) attestation protocols. 
The reports are aggregated
and received by the edge verifiers, which in turn, aggregate them
with their own report and forward them to the upper
layer. Finally, a set of verifiers (potentially singleton) receive
the final %(aggregated) 
reports from the cluster heads. Note
that CRA protocols %protocols 
may leverage a mixture of topologies, e.g.,
\shela~\cite{shela} may utilise a different attestation protocol within
each of its clusters and each of those protocols may work with a
different topology. % , thus resulting in hierarchy of different

\item [Publish/subscribe (\TopPubSub)] protocols allow devices to take
the role of a publisher or subscriber of \emph{topics}, and
asynchronous communication to take place through an intermediate
broker. Publishers send messages tagged with a topic to the broker,
and the broker forwards these to all subscribers on that topic.
%\sa{how to polish this?}
% \bd{expand the definition}
% An example attestation protocol working with
% a swarm with {\it publish/subscribe} topology is \sara~\cite{sara}
% where \sa{complete the explanation}

\item [Ring] protocols allow devices to be logically linked to their
predecessors and successors, thus creating a ring.

\end{description}

\begin{table}[!t]
\scriptsize
\begin{center}
\centering\caption{\label{table:categories-OLD} Topology, Dynamicity, and Centralisation}
\begin{tabular}{|p{0.3\linewidth}|p{0.17\linewidth}p{0.21\linewidth}p{0.17\linewidth}|}
  \hline 
  \rowcolor{DarkGray!70!white}
  {\bf Protocol} & {\bf Topology} (\cref{subsec:topology}) & {\bf Dynamicity} (\cref{subsec:dynamicity}) &{\bf Centralisation} (\cref{subsec:centralised})  \\
  \hline 
  \seda\cite{seda} & \TopST & \DynStatic & Centralised \\
%\hline
\rowcolor{Gray}
\darpa\cite{darpa}+\seda\cite{seda}&  \TopST & \DynStatic %dynamic 
& Centralised  \\
%\hline
\sana\cite{sana}&  \TopAgg & \DynStatic %dynamic 
& Centralised  \\
%\hline
\rowcolor{Gray}
\lisas\cite{lisa}&  \TopST & \DynQStatic &Centralised \\
%\hline
\lisaalpha\cite{lisa}&\TopST & \DynQStatic &Centralised\\
%\hline
\rowcolor{Gray} 
\seed\cite{seed}+\seda\cite{seda}& % dependant on \seda = spanning-tree
(\TopST) & \DynStatic %dynamic 
& Centralised \\ %DoS-V
%\hline
\scapi\cite{scapi} & \TopST & % static/dynamic 
\DynHDynamic &Centralised \\
%\hline
\rowcolor{Gray} 
\erasmus\cite{erasmus}+\lisaalpha\cite{lisa}& % dependant on \lisaalpha = spanning-tree
(\TopST)
& \DynHDynamic &Centralised \\ %DoS-V
%\hline 
\erasmusod\cite{erasmus(extended)}+\lisaalpha\cite{lisa}& (\TopST)% dependant on \lisaalpha = spanning-tree
& \DynHDynamic &Centralised \\ %DoS-V
%\hline
\rowcolor{Gray}
\consensus\cite{cons} & \TopBroad & \DynHDynamic &Decentralised\\ %DoS-V
%\hline 
\pads\cite{pads} & \TopBroad & \DynHDynamic &Decentralised \\ %DoS-V
%\hline 
\rowcolor{Gray}
\salad\cite{salad} & \TopBroad & \DynHDynamic & Decentralised\\ %DoS-V
%\hline
\wise\cite{wise2018} & \TopST & % static / 
\DynQStatic & Centralised \\
%\hline
\rowcolor{Gray} 
\slimiot\cite{slimiot2018}& \TopST& % static/dynamic 
\DynHDynamic & Centralised \\
%\hline 
  \usaid\cite{us-aid}& \TopBroad & \DynDynamic &Decentralised\\
%\hline 
%\attauth is based on swatt so it is PM-R
\rowcolor{Gray}
\attauth\cite{attauth} & \TopST &---&Centralised \\
%\hline 
\mvs-D2D\cite{mvs} & Distributed & --- &Decentralised \\
\rowcolor{Gray}
\mvs-Consensus\cite{mvs} & Cluster & --- &Decentralised \\
\museda\cite{museda} & \TopST & \DynHDynamic % (to some extent) = 
& Centralised \\
%\hline
\rowcolor{Gray}
\sap\cite{sap}& \TopBBT% balanced binary tree
& \DynStatic & Centralised \\ 
%\hline 
%DoS-v
\mtra\cite{mtra} & \TopCluster % + fixed tree topology
& \DynStatic & Decentralised \\ %Page 2 of ESDRA claims it is hw-based?
%\hline 
%\radis uses symmetric keys to mitigate DoS 
\rowcolor{Gray}
\radis\cite{radis} & \TopBroad & \DynStatic & Centralised \\
%\hline 
%I --> accusation report ? or it is List ? based on the accused ones, it considers others healthy.
\esdra\cite{esdra} & \TopCluster & \DynDynamic & Decentralised \\
%\hline 
%Eapa assumes a reliable attestation protocol to attest the s/w integrity of provers and only focuses on physical attack. In fact, most of the protocol describes the heart-beat protocol scenario. However, Like DARPA, SCAPI, or US-AID I have decided to consider to assume the heart-beat report as the attestation report so that I can adapt the security properties.
%I --> accusation report ? or it is L ? based on the accused ones, it considers others healthy.
\rowcolor{Gray} 
\eapa\cite{eapa} & \TopBroad & --- &Decentralised\\
% \hline
\shela\cite{shela} & \TopHierarchical % graph
& % static/dynamic/ 
\DynHDynamic & Decentralised \\
% \hline 
\rowcolor{Gray}
\healed\cite{healed} &--- & --- & Decentralised \\
% \hline
%am not sure about DoS in SADAN
\sadan\cite{sadan}+\diat\cite{diat} & \TopBroad & % static/
\DynStatic & Decentralised\\
% \hline
%in SFS, the third party (cloud server) who does the main computations is untrusted, but Verifier is trusted
\rowcolor{Gray}
\sfs\cite{sfs} % \ \ (*)
                 & \TopST & --- & Centralised\\
% \hline
\pasta\cite{pasta} & \TopST & \DynHDynamic % (autonomous)
& Decentralised \\
% \hline 
\rowcolor{Gray}
\dads\cite{dads} & \TopST & \DynHDynamic & Decentralised\\
% \hline 
\simpleplus~\cite{simple} & \TopST & % static/
\DynStatic & Centralised\\
% \hline 
\rowcolor{Gray}
\cora\cite{cora}& \TopST & --- &Centralised\\
% \hline
\sara\cite{sara} & \TopPubSub & \DynStatic &Centralised\\
\rowcolor{Gray}
\fadia\cite{fadia} % \jlp{I think DA should be ticked for this}
                 & \TopST & \DynStatic+ join/leave &Centralised\\
% \hline
\bfb\cite{bfb} % \ \ (*)
                 &  \TopBroad & \DynHDynamic &Decentralised\\
% \hline 
\rowcolor{Gray} 
\mton\cite{m2n} % \ \ (*)
                 & \TopCluster & --- &Decentralised\\
% \hline
\swarna\cite{swarna} & \TopST & \DynStatic & Centralised \\
% \hline  
\rowcolor{Gray}
\shots\cite{shots} &  \TopST & \DynStatic&Centralised \\
% \hline
\hola\cite{hola} &  Ring & \DynDynamic & Decentralised\\
% \hline
\rowcolor{Gray}
\fesa\cite{fesa} &  Hierarchy &\DynHDynamic &Decentralised \\
% \hline
\scraps\cite{scraps} &  \TopPubSub & Dynamic & Decentralised\\
\hline
\end{tabular}  
\end{center}
\end{table}

Spanning-tree is the most widely used topology in the CRA protocols and is a good fit in settings where
devices do not have enough memory to store the
reports of other devices but can aggregate and forward them
%(e.g., \seda~\cite{seda}, \lisa~\cite{lisa}). 
(e.g., \lisa~\cite{lisa}). 
Such CRA protocols typically require a reliable network where the
devices maintain their connections with other devices, and are
effective in a centralised settings. %(see~\cref{subsec:centralised}).
However, they perform poorly under device mobility,
since the tree topology may be hard to maintain.

Aggregation layer CRA protocols allow untrusted aggregators as in the case
of untrusted routers or cloud servers (e.g., \sana~\cite{sana}).
In interactive CRA protocols %(see \cref{subsec:interactive} below)
attestation requests have to travel down the leaf and reports back up
to the root. To this end, \emph{balanced binary tree} CRA protocols have
been suggested (e.g., \sap~\cite{sap}) to minimise the furthest path
from the root verifier to a leaf prover, and hence minimise the
communication
overhead. % , since each device may communicate with up to two children
% and one parent.

Ensuring resilience  is also
challenging for spanning trees since the loss or delay of a device
affects its entire subtree. % of a device
% or a disruptive or delayed network (4) 
An initiator is particularly susceptible to DoS attacks via fake
attestation requests generated by an adversary. % to deviate swarm
% devices from their main tasks. 
Spanning-trees are not recommended in particular for CRA protocols that
aim to detect physical attacks via a heartbeat protocol, since if each prover sends out a heartbeat to
its neighbouring provers and receives from them an accumulated heartbeat, the number of exchanged messages per periodically executed
heartbeat period scales quadratically with the number of provers~\cite{darpa}, which causes scalability issues in large
swarms. % Moreover, maintaining the connections of the spanning tree
% for transmitting all those messages will not be possible in practice;
% thus, a
A single failure in the transmission of a heartbeat suffices to cause
a false positive, where a healthy device is mistakenly regarded as
unhealthy~\cite{scapi}.

% Bad for protocols who utilises
% absence/detection scheme to determine captured devices as spanning-tree
% may be hard to maintain and false positive cases increase (5) Bad for
% centralised attestation protocols where the verifier is not trusted

Distributed topologies are good fit for swarms with mobility and are
resilient to devices (re)joining or leaving the swarm, i.e., swarm can
be scaled as required (e.g., \salad~\cite{salad}).
%, \pads~\cite{pads}, \usaid~\cite{us-aid}, \radis~\cite{radis}, \eapa~\cite{eapa}, \sadan~\cite{sadan}). 
They also provide resilience against device
failure (e.g,. \scapi~\cite{scapi} uses absence detection scheme to identify physically
captured provers). But they do not require initiators to be trusted
(e.g., \mvs~\cite{mvs}). %\cite{salad} % (5) Resilient against
% failure of a single device, since other devices can still communicate
% with each other. (6) Good for centralised or decentralised protocols
% with either trusted or untrusted verifier(s).
%such 
Distributed protocols are much more difficult to design since it can
be difficult to ensure correct synchronisation and timing, which may
introduce additional communication overhead. % \bd{Finish this: has the
% following shortcomings: %(1) lack of global knowledge
% (1) Bad for compatibility} \sa{I could not find anything for this one.}

Cluster CRA protocols allow provers with greater memory and/or computational power act as cluster heads to perform more complicated tasks (e.g., \mtra~\cite{mtra}), however, if cluster heads need to communicate with each other, there
is an increased cost of maintaining synchrony across them.

Hierarchical CRA protocols (e.g., \shela~\cite{shela}) partition swarms
into sub-clusters % extending the
% edge layer with additional higher-end edge devices, or adding
% additional edge layers to the topology, in which each layer attests
% the devices in the lower-level layer in the hierarchy. This way each
% sub-swarms can be attested via different attestation schemes (6)
% Hierarchy
and are effective when edge verifiers have more computational
resources than other devices. These protocols let a swarm be geographically widespread,
with a single cluster head attesting devices locally. However, there
is an increased cost of maintaining synchronicity across cluster heads.

% (7) Hierarchy
% is good for heterogenous devices where devices in each cluster may
% have different resources and features. (6) Hierarchy protocols may
% increase the cost as the verifier as it may use a synchronisation
% mechanism to communicate with a set of edge verifiers.

In publish/subscribe CRA protocols, while the use of a broker
potentially may reduce network communication (since subscribers will only
be forwarded relevant messages), the broker itself becomes a potential point of attack~\cite{scraps}. A solution to this is decentralised publish/subscribe topology %(with or without brokers) 
that guarantees events ordering among operations performed by devices that collaborate to deliver a distributed service~\cite{sara}.

Ring CRA protocols are efficient where there is no centralised verifier (see \cref{subsec:centralised}) and each device is attested by its predecessor/successor device in the ring, otherwise the attestation results of all the devices must travel through other devices in the ring to reach the verifier; thus, if one device fails, the entire network is impacted. Ring protocols can be easily managed when a prover wants to (re)join.

\subsection{Dynamicity}
\label{subsec:dynamicity}
The structure of connections between %swarm 
devices may vary as they
may be disconnected and (re)join %either 
due to network effects, device
mobility or malicious actors. In this paper, we use the term \emph{dynamicity} 
to refer to the level of change in swarm topology that can be tolerated while 
attestation is taking place and not between consecutive attestation procedures. %(aka attestation rounds). 
CRA protocols often support dynamicity between two consecutive attestation procedures, however, some of them support dynamicity \emph{during} attestation procedures
(e.g., %\scapi~\cite{scapi}, \pads~\cite{pads}, \shela~\cite{shela},\pasta~\cite{pasta}, 
\dads~\cite{dads}), and some not 
(e.g., %\mtra~\cite{mtra}, \radis~\cite{radis}, \swarna~\cite{swarna}, \fadia~\cite{fadia}, 
\sara~\cite{sara}). In the former case, the
underlying assumption is that the connections between devices remain
static, and there is no device leaving or (re)joining the swarm. We
classify dynamicity as follows (showin in Column 3 of \cref{table:categories-OLD}).

\begin{description}[itemsep=0pt,
leftmargin=0pt]
\item [Static] protocols, where the topology does not change
during an attestation procedure.
\item [Quasi-static (\DynQStatic)] protocols, where the topology may change
during an attestation procedure, as long as changes do not influence
message exchange between devices. For example, a link may disconnect
after a device finish sending a message and reconnect before any
other message is exchanged between the same pair of
devices. 

\item [Dynamic] protocols, where the swarm may change during
an attestation procedure in a limited way. 
A device may move gradually and change its set of %direct 
neighbouring devices, but never
disappears from one neighbourhood and reappears
in a remote neighbourhood at once.
Moreover,
each device is always %available and 
reachable in the swarm. 
%i.e., it cannot be completely disconnected and join.

\item [Highly Dynamic (\DynHDynamic)] protocols, where the topology may change
unpredictably during an attestation procedure. Devices may be
disconnected, (re)join, and freely move around to establish new
connections with new neighbouring devices.
%An example of a swarm attestation protocol working with dynamic topology is \shela~\cite{shela}, where through built-in redundancy, it allows the swarm provers to be temporarily unavailable to one or more edge devices. Therefore, the provers do not have to be static even during attestation. 
%Another swarm attestation protocol that support highly dynamic swarm topologies is \pads~\cite{pads} which lets device mobility even during attestation, but it does not guarantee a full coverage. However, the coverage grows as the number of interactions between the swarm devices increase. 
%Swarm attestation protocols that work with dynamic or highly dynamic topologies must be able to efficiently address key management, network discovery, routing, and setting an
%upper bound on the communication delay between any two
%devices in a secure way.
\end{description}

%PADS --> self-attestation: mentioned in shela paper

Static CRA protocols are often simpler to design, e.g., regarding aspects
such as key management among adjacent devices, and are a good fit when devices are not mobile. However, they often
require that the underlying network is reliable (e.g.,
\seda~\cite{seda}).
Quasi-static CRA protocols inherit the pros/cons of static CRA protocols but
are a good fit for networks where links may be disconnected then
reconnected before sending a subsequent message (e.g.,
\lisa~\cite{lisa}). Dynamic CRA protocols are useful when devices have
continuous reachability, but may increase the communication costs, and should not be used when devices may be completely
disconnected (e.g., autonomous swarms as in \esdra~\cite{esdra}).
Highly dynamic CRA protocols can cope with unreliable and disruptive
networks and support device mobility (e.g., swarms of drones as in
\usaid~\cite{us-aid}). In both dynamic and highly dynamic CRA protocols,
key management and routing mechanism become much more sophisticated.

\subsection{Centralisation}
%\subsection{(De)Centralised CRA protocols}
\label{subsec:centralised}

{\em Centralised} CRA protocols usually accumulate the final aggregated report 
in an {\em initial prover} that directly exchanges messages with
a verifier. The verifier requests the report from the initial
prover 
%(e.g., \attauth~\cite{attauth}, \wise~\cite{wise2018}). 
(e.g., \attauth~\cite{attauth}). There are also {\em decentralised}
CRA protocols %protocols
where the final (aggregated) report(s) is recorded by more than one
device (possibly all devices) in the swarm and a verifier can query
that report from any of them (see column 4 in
\cref{table:categories-OLD}). Decentralised CRA protocols may be fully distributed,
where untrusted neighbouring provers attest each other and share their
reports with other provers, and provers use consensus algorithms to
come to an overall conclusion about the swarm's status
%(e.g., \pads~\cite{pads}, \consensus\cite{cons}). 
(e.g., \pads~\cite{pads}).  CRA protocols with cluster or hierarchical topologies are
also decentralised as a verifier queries the status of the provers in
each cluster from the corresponding cluster head, which could be
either an edge verifier (e.g., \shela~\cite{shela}), or a prover with
enough computation resources, which aggregates reports received from
the provers in the corresponding cluster and will itself be attested
later by the verifier (e.g., \mtra~\cite{mtra}).

\input{app_model_description.tex}

%%% Local Variables:
%%% mode: latex
%%% TeX-master: "main"
%%% End:

%% file: app_model_description.tex
\section{Further details of \tamarin encoding}\label{app:tamarin}
% \jlp{I have done my best to explain the \simpleplus model as completely as I can without taking up too much space. The only part of this section I definitely want to keep in some form is the group attestation lemma, as it is a novel contribution of this SoK}.

Here we give additional detail on our \tamarin work.
%  the analysis of \simpleplus. 
In~\cref{subsec:group-att} we describe our lemma for Asynchronous Group Attestation. For completeness, in~\cref{subsec:simpleplus-model} we describe Le-Papin et al.'s approach to modelling \simpleplus. 

\subsection{Asynchronous Group Attestation}\label{subsec:group-att}
While \simpleplus is an L-QoSA (See~\cref{sec:attest-prop-1}) protocol, we can trivially produce a group result by AND-ing all the individual results. This allows us to verify Group Asynchronous Weak/Strong Attestation, shown in~\cref{fig:async-group-att-strong}. 

$prvs$ is the set of provers that $att$ is a result for, referred to as $G$ in~\cref{def:group (a)synchronous weak/strong attestation}. However, as \simpleplus does not give the Relying Party information about which provers contributed to a report, it has no choice but to assume that all provers did, causing it to fail Group Asynchronous Strong Attestation.

\begin{figure}[t]
  \begin{lstlisting}
All V prvs att cntr #att_end . 
Attestation_Complete_G(V, prvs, att, cntr) @ #att_end ==> 
(
Ex #att_start . Attestation_Start(V, cntr) @ #att_start 
&
(
  // All provers in prvs are healthy at #att_start
  att = healthy ==> not(
    // case: |prvs| > 1
    (Ex P0 z . (P0 ++ z = prvs) 
    & IsNotHealthy(P0, #att_start))
    |
    // case: |prvs| = 1
    (Ex P0 . (P0 = prvs) & IsNotHealthy(P0, #att_start))
  )
)
&
( //Strong
  // At least one prover is unhealthy by #att_end
  att = unhealthy ==> ( 
    // case: |prvs| > 1
    (Ex P0 z . (P0 ++ z = prvs) 
    & IsNotHealthy(P0, #att_end))
    |
    // case: |prvs| = 1
    (Ex P0 . (P0 = prvs) & IsNotHealthy(P0, #att_end))
  )
)
)
  \end{lstlisting}
  \caption{Lemma for Asynchronous Group Strong Attestation. Asynchronous Group Weak Attestation is similar, omitting the $att = unhealthy$ case.}
  \label{fig:async-group-att-strong}
\end{figure}

% \begin{figure}[t]
%   \begin{lstlisting}
% IsHealthy(P, #time) <=> not(
%   // A malware infection of P
%   Ex #infection . (
%     MalwareInfection(P) @ #infection & 
%     // Before #time
%     #infection < #time 
%   )
% )
%   \end{lstlisting}
%   \caption{\ishealthy states that a prover $P$ is healthy at time $\#time$ iff $P$ has not been infected with malware in the past. }
% \end{figure}

% \begin{figure}[t]
%   \begin{lstlisting}
% IsNotHealthy(P, #time) <=> (
%   // A malware infection of P
%   Ex #infection . (
%     MalwareInfection(P) @ #infection & 
%     // Before #time
%     #infection < #time 
%   )
% )
%   \end{lstlisting}
%   \caption{\isnothealthy states that a prover $P$ is unhealthy at time $\#time$ iff $P$ has either been infected with malware in the past, $P$ is infected with malware at $\#time$.}
% \end{figure}

\subsection{\simpleplus Model}\label{subsec:simpleplus-model}
% \paragraph{Adversary Model} 
% \jlp{This first paragraph was originally in the main body of the SoK.}
Le-Papin \etal's model reflects assumptions, common in CRA protocols, that the protocol code execution is atomic (a prover cannot become compromised in the middle of the attestation procedure) and has controlled invocation (a prover never starts from the middle of the attestation procedure). The untrusted environment is modelled as follows. The state of the attestation target is represented by a constant (with the prover checking that this constant matches the one sent by the verifier). Software compromise of the prover is modelled by the state being replaced with a different constant. The remaining capabilities of a software adversary are subsumed by \advc. 

% \paragraph{Key Reveals}
Le-Papin et al. consider a threat model where some provers can reveal their secrets to an attacker. The threat models we consider for \simpleplus, \advsw and \advc, do not permit this. 

% \subsubsection{Bitwise operations}\label{subsubsec:bit-ops}
\simpleplus uses bitwise OR to aggregate reports. Le-Papin et al. faithfully models OR, with $onee/0$ and $zeroo/0$ functions representing '1' and '0'. Then, the $or/2$ function works with these values as one would expect. In~\cref{fig:bitwise-ops} we show this and our $AND/2$ function and equations, as used in~\cref{subsec:group-att}, which work in the same way.

\begin{figure}[t]
    \begin{lstlisting}
functions: or/2, and/2, onee/0, zeroo/0

equations: or(onee, onee) = onee,
            or(onee, zeroo) = onee,
            or(zeroo, onee) = onee,
            or(zeroo, zeroo) = zeroo,

            and(onee, onee) = onee,
            and(onee, zeroo) = zeroo,
            and(zeroo, onee) = zeroo,
            and(zeroo, zeroo) = zeroo
    \end{lstlisting}
    \caption{Functions and equations for modelling binary OR and AND.}
    \label{fig:bitwise-ops}
\end{figure}

% \paragraph{Report}
The bitvector report is modelled as a pair $<attest_0, attest_1>$, with each value in the pair being either $onee$ or $zeroo$, representing one bit in the bitvector.
% \paragraph{Topology}
Each prover has an id, represented as the constants $'0'$ and $'1'$. Each id represents a subtly different role, rather than all provers being the same. This is because they each create different initial reports $<attest, zeroo>$ and $<zeroo, attest>$, respectively.
% \paragraph{Bounding}
For tractability, $'0'$ is the only prover able to aggregate reports. This represents a restriction of the possible topologies, with the variations being $V-P0-P1$, $V-P0$, and $V-P1$. Furthermore, the model is analysed with no more than two rounds. 
% Modelling cntr
As the model was written before \tamarin provided support for natural numbers, the counter is modelled as a multiset ($'1' = 0$, $'1'++'1' = 1$, etc.).

Our method has the advantage of directly proving the main security
goal of a CRA protocol, as opposed to a %collection
set of properties\footnote{Le Papin \etal~refer to these as Prover
Authentication, Legitimate Response, and Result Timeliness.} that
have not yet been proven to yield the primary security goal. However,
this comes at a cost of taking longer to verify in \tamarin (14.1
hours for %our 
Individual Attestation lemma versus a combined 4.6 hours
for Le-Papin \etal's analogous properties on the same hardware).

% \paragraph{Updating Kcol}
% The attest phase for the prover has two initial rules, one for if the prover is healthy and one for if the prover is unhealthy, each producing different results. Two more rules are then used to update the secret 
% \paragraph{timeout}

%%% Local Variables:
%%% mode: latex
%%% TeX-master: "main"
%%% End: